%% file: sample-acmlarge.tex
\documentclass[acmlarge]{acmart}

\usepackage{times}
\usepackage{booktabs}
\usepackage{tabularx}
\usepackage{tabu}
\usepackage{multirow}
\usepackage{graphicx}
\usepackage{makecell}
\usepackage{subfigure}
\usepackage{amsmath}
\usepackage{bbding}
\usepackage{pifont}
\usepackage{wasysym}
\usepackage{amssymb}
\usepackage{lscape}
\usepackage{adjustbox}
\usepackage{framed}
\usepackage{color}
\usepackage{soul}
\AtBeginDocument{%
  \providecommand\BibTeX{{%
    \normalfont B\kern-0.5em{\scshape i\kern-0.25em b}\kern-0.8em\TeX}}}

\setcopyright{none}
\begin{document}

\title{Predictive Models in Software Engineering: Challenges and Opportunities}


\author{Yanming Yang~}
\affiliation{Zhejiang University, China, and Monash University,  Australia}
 \email{Yanming.Yang@monash.edu}

\author{Xin Xia~}
\affiliation{Monash University Australia}
\email{Xin.Xia@monash.edu}

\author{David Lo~}
\affiliation{Singapore Management University, Singapore}
 \email{davidlo@smu.edu.sg}

\author{Tingting Bi~}
\affiliation{Monash University,  Australia}
  \email{Tingting.Bi@monash.edu}

\author{John Grundy~}
\affiliation{Monash University,  Australia}
	\email{John.Grundy@monash.edu}

\author{Xiaohu Yang~}
\affiliation{Zhejiang University, China}
\email{yangxh@zju.edu.cn}

\renewcommand{\shortauthors}{Yang et al.}

\begin{abstract}
 Predictive models are one of the most important techniques that are widely applied in many areas of software engineering. There have been a large number of primary studies that apply predictive models and that present well-preformed studies and well-desigeworks in various research domains, including software requirements, software design and development, testing and debugging and software maintenance. This paper is a first attempt to systematically organize knowledge in this area by surveying a body of 139 papers on predictive models. We describe the key models and approaches used, classify the different models, summarize the range of key application areas, and analyze research results. Based on our findings, we also propose a set of current challenges that still need to be addressed in future work and provide a proposed research road map for these opportunities.
\end{abstract}



\maketitle

\section{Introduction}
Researchers have developed automated methodologies to improve software engineering tasks. Key reasons are usually to save developer time and effort and to improve the software quality in terms of stability, reliability, and security. Many of such studies have resulted in great improvements in various tasks \cite{bourque2014guide, rehman2018roles, zhang2018predictive, tantithamthavorn2016automated, lin2019identifying}.

A key technology, the \emph{predictive model}, has been developed to solve a range of software engineering problems over several decades. The use of predictive models is in fact becoming increasingly popular in a wide range of software engineering research areas. Predictive models are built based on different types of datasets -- such as software requirements, APIs, bug reports, source code and run-time data -- and provide a final output according to distinct features found in the data. There are various predictive models commonly used in software engineering tasks that contribute to improving the efficiency of development processes and software quality. Common ones include defect prediction \cite{tantithamthavorn2016automated}, API issue classification \cite{lin2019pattern}, and code smell detection \cite{palomba2018beyond}.

Despite numerous studies on predictive models in software engineering, to the best of our knowledge, there has been no systematic study to analyze the use of and demonstrated the potential value of and current challenges of using predictive models in software engineering. There is no clear answer as to which software engineering tasks predictive models can be best applied and how to best go about leveraging the right predictive models for these tasks. Answering these questions would be beneficial for both practitioners and researchers, in order to make informed decisions to solve a problem or conduct research using predictive models. %Researchers are able to improve existing predictive models based on the adoption challenges they face and the expectations of practitioners.

This paper contributes to the research on predictive models by performing a comprehensive systematic survey of the domain. At first we recorded all titles of papers published in top software engineering venues, such as ICSE, ASE, FSE, TSE, TOSEM, and EMSE from 2009 -- 2019. We performed searches on the title and abstract of each paper and collected over 2,000 papers containing our target search terms. We then carefully read the abstract and introduction of each paper to validate its relevance to this study. Finally, we identified 145 high-quality primary study papers. We found that predictive models have been widely applied to a variety of software engineering tasks. Common ones include requirements classification, code change detection and malware detection. We grouped predictive model usage in software engineering into six research domains according to the software development lifecycle (SDLC) -- software requirements, software design and development, testing and debugging, software maintenance, and software professional practice knowledge area. We further summarized and analyzed these studies in each domain. After that, we identified a set of remaining research and practice challenges and new research directions that we think should be investigated in the future. To the best of our knowledge, we are the first to perform such a systematic review on the use of predictive models in software engineering domain.

This paper makes the following key contributions:

\begin{itemize}
\item We present a comprehensive survey on predictive models covering 145 primary study papers from the last decade;
\item We analyze these 145 primary studies and characterize them, and identify 10 commonly used measures for model evaluation;
\item We summarize the research domains in which the predictive model has been applied, and discuss distinct technical challenges of using predictive models in software engineering; and
\item We outline key future avenues for research on predictive models in software engineering.
\end{itemize}

The remainder of this paper is organized as follows. Section 2 briefly introduces the workflow of predictive models. Section 3 presents our study methodology. Section 4 investigates the evolution and distribution of the selected primary studies using predictive models for software engineering tasks, and Section 5  gives a classification and research distribution of these predictive models. Section 6 summarizes and analyzes the application of predictive models in software engineering, and a discussion of the threats that could affect the validity of our findings is presented in Section 7. Section 8 discusses the challenges that still need to be addressed in future work and give outlines a research road map of potential research opportunities in this domain.  Section 9 provides a summary of the key conclusions of this work.

\input{An_Introduction_of_Predictive_Model}

\input{Methodology}

\input{RQ1}

\input{RQ2}

\input{RQ3}

% JG: Put threats BEFORE the research roadmap is best IMO...

\input{Threats_to_Validity}

\input{Challenges_and_opportunities}

\input{Conclusion}


%\begin{acks}
%To Robert, for the bagels and explaining CMYK and color spaces.
%\end{acks}

\bibliographystyle{ACM-Reference-Format}
\bibliography{predictive_model}


\end{document}

%% file: An_Introduction_of_Predictive_Model.tex
\section{Predictive Models}

\subsection{Workflow of predictive models}

\begin{figure}[htbp]
\centering
%\hspace{1in}
\includegraphics[width=0.5\textwidth]{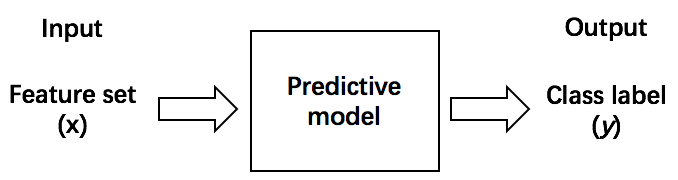}
\caption{Predictive model as the task of mapping an input feature set $x$ into its class label $y$.}
\label{fig.predictive model}
\end{figure}

A predictive model is usually treated as a black box that automatically assigns a class label when presented with the feature set of an unknown record, illustrated in Fig. \ref{fig.predictive model}. Instances in the input dataset can be assigned to one of several predefined categories by building such a predictive model. The predictive model is usually trained with a representative input dataset, then applied to target input datasets. The workflow of using a predictive model can be described as a mathematical problem of learning a target function ($f$) that maps each feature set ($x$) in a dataset ($X$) to one class label ($y$). The job of predictive models is to find the best target function ($f_{m}$).

There are four critical components when building such a predictive model:

\textbf{\textit{Datasets:}}
As the most basic component of predictive models, the dataset has a large impact on a model's performance. Low-quality datasets with noise and mislabeling  may lead predictive models to provide (very) wrong experimental results, even if the process of model selection and training is effective.

Different types of datasets are used when performing different software engineering tasks with predictive models. For instance,  studies may use source code, bug reports, or requirement documents as datasets of predictive models for defect prediction, bug report classification, and requirements-related knowledge classification respectively. In general, different datasets have different properties. These include scale, distribution, bias, quality, representativeness, sparseness and so on.

\textbf{\textit{Features:}}
In building predictive models, features (or attributes) in datasets are essential  in the model training phase. The goal of a predictive model is essentially to learn a target function by analyzing different feature sets in given input datasets. Thus a good feature set can allow predictive models to learn the potential patterns in datasets and thus to output correct labels effectively. In order to construct high-quality feature sets, feature selection is an important optimization technique in building predictive models. Better feature selection strategies can improve the accuracy of results, reduce overfitting, as well as reduce training time.

\textbf{\textit{Model Building Algorithms:}}
A predictive model can be implemented by using different algorithms. For instance, J48, C4.5 and CART are commonly used algorithms for building Decision Tree-based predictive models. A number of relatively new deep learning-based algorithms and architectures, such as RNN and CNN-based networks, have been applied as predictive models to many software engineering tasks. Some predictive model algorithms have been introduced for specific software engineering problems, by enhancing existing algorithms, or providing new model capabilities. For example, many large and complex software-intensive systems widely used logs for troubleshooting, and thus Zhang et al.~\cite{zhang2019robust} present a tool, called LogRobust, to detect system anomalies by analyzing log messages. They adopted Bi-LSTM, a variant of LSTM model, to capture the contextual information in log sequences, and combined attention mechanism to strengthen the ability of automatically learning the importance of different log events. Liu et al. \cite{liu2018deep} concentrated on detecting feature envy, one of the most common code smells, and tailored a deep learning model to consider both textual input (method name or class name) and numerical input (the distance between a method and a class) by merging two CNN models. %This allows researchers to select a better one from a range of algorithms to address target software engineering task concerns.

\textbf{\textit{Model Performance Measures:}}
Many widely-used evaluation metrics, such as recall, precision, F-measure, accuracy, and AUC, have been introduced and applied in software engineering studies. This is because different tasks may require different metrics to evaluate the effectiveness of their proposed predictive models. On the other hand, some of the less common evaluation metrics, such as Balance, G-measure, and specificity, also appear in a few studies. We will present an analysis of various evaluation metrics for predictive models in Section 5.

%% file: Methodology.tex
\section{Methodology}

To perform a comprehensive systematic review of the use of predictive models in software engineering, we followed the systematic review guidelines provided by Kitchenham and Charters \cite{keele2007guidelines} and Petersen et al. \cite{petersen2015guidelines}.

\begin{table*}[!htbp]
\centering
\scriptsize
\caption{Research Questions and Motivations}
\label{tab:RQs}
\begin{tabular}{c p{4cm} p{10cm}}
%\hline
\toprule
\textbf{RQs} & \textbf{Research Question} & \textbf{Motivation}
\\
  \midrule
RQ1& What are the trends in the primary studies on use of predictive models in SE?
  &
The basic information (e.g., authors, publication year, affiliations) of the primary study papers can be informative. We wanted to understand where primary study papers on SE uses of predictive models are being published and any trends that can be observed. The goal of this RQ is to investigate these publication trends and distribution of the primary studies.
\\
RQ2& Which predictive models are applied to what software engineering tasks?
  &
Various predictive models are used to in the context of different software engineering tasks. It is necessary to generalize about predictive models so that researchers can select the appropriate model when researching in certain domains. The goal of this question is to determine which predictive models are frequently applied in software engineering, and to develop a classification for them.
\\
RQ3& In what software engineering domains and applications have predictive models been applied?
  &
Although predictive models have been applied to broad application scenarios in software engineering, there does not exist any comprehensive study that summarizes these research domains and applications using different models. This RQ aims to find and report on such a domain analysis.
\\
\bottomrule
\end{tabular}
\end{table*}

\subsection{Research Questions}

The aim of this paper is to summarize, classify, analyze and propose research directions based on empirical evidence concerning the different modeling techniques and task domains involved. To achieve this, we define three research questions as shown in Table \ref{tab:RQs} and give the motivation behind each question.

RQ1 will analyze the distribution of publications on predictive models over the last decade to give an overview of the trend in software engineering research. RQ2 will provide a  classification and distribution of predictive models used in software engineering and identify what are the commonly used evaluation metrics. RQ3 will explore where and how predictive models have been applied for specific software engineering tasks.

\subsection{Literature Search and Selection}

We identified a set of search terms based on software engineering tasks involved in predictive models that were already known to us. We then refined these by checking the title and abstract of the relevant papers. After considering the alternative spellings and synonyms for these search terms \cite{hosseini2017systematic}, they were combined with logical ORs, forming the complete search string. The search terms are listed as follows:

\textit{
\footnotesize
("predict*" OR "identif*" OR "detect*" OR "classif*" OR "infer*" OR "recommend*" OR "indicat*" OR "estimat*")
Since the Digital Library can support search terms that are not case sensitive and not "whole words only", we use the stemmed term to represent all forms related to them (e.g., “classif*” can search for “classify”, “classifying”, “classified” as well as “classification”).
}

Following previous survey study approaches \cite{huang2019survey, hosseini2017systematic}, we first collected the titles of all papers published at ICSE, ASE, FSE, TSE, TOSEM, and EMSE between 2009 and 2019 from DBLP computer science bibliography. We then used our search string to perform search for papers related to predictive models, downloaded relevant papers from IEEExplore, ACM Digital Library, ISI Web of Science, and Google Scholar. We obtained 2,128 candidate primary studies; the initial number of studies for each search engine is shown in Table \ref{tab:study number}. After discarding  duplicate papers, we applied inclusion/exclusion criteria (see Section 2.3) to the title, abstract and keywords of each paper for narrowing the candidate set size to 521 studies. We then validated their relevance by reading these studies in full and found that a final 145 fulfilled all criteria.

\begin{table}[htbp]
\centering\scriptsize
\caption{The initial number of studies for different search engines}
\label{tab:study number}
 \begin{tabular}{p{5.5cm}p{1.5cm}<{\centering}}
  \toprule
 \textbf{Search Engine} & \textbf{\#Studies} \\
  \midrule
 ACM Digital Library         & 402  \\
 IEEE Xplore Digital Library & 273 \\
 ISI Web of Science          & 628 \\
 Google Scholar              & 636 \\
 DBLP                        & 189 \\
 \textbf{Total}              & \textbf{2128} \\
  \bottomrule
 \end{tabular}
\end{table}

\subsection{Inclusion and Exclusion Criteria}

A number of low quality studies (e.g., non-published manuscripts and grey literature) are found when searching for the relevant papers with search engines. We also wanted only the most comprehensive or latest version of repeated studies in the candidate set. For example, we removed a conference paper if it was extended in a follow-on journal paper. Thus we applied a set of inclusion and exclusion criteria to select the studies with strong relevance.

The following inclusion and exclusion criteria were used:

\ding{52} \ The paper must be written in English.

\ding{52} \ The paper must apply one or more predictive models to address software engineering tasks.

\ding{52} \ The paper must be a peer reviewed full research paper published in a conference proceedings or a journal.

\ding{56} \ Papers less than six pages are not considers.

\ding{56} \ The predictive models used in studies cannot be baseline approaches.

\ding{56} \ Conference version of a study that has an extended journal version is not considered.

\ding{56} \ Keynote records and publications in grey literature are dropped.

The inclusion and exclusion criteria were piloted and applied by the first and fourth authors by assessing 45 randomly selected papers from the initial set. We used Cohen's kappa coefficient to measure the reliability of the inclusion and exclusion decisions. The agreement rate in the pilot study was "substantial" (0.72). We also perform the assessment in the full list of identified papers, and Cohen’s kappa score was "substantial" (0.68). The first and fourth authors reached a consensus through discussions when encountering disagreements. In case disagreements were not resolved, a third researcher  who is not the author of the paper was invited to make the final decision.

\subsection{Data Extraction and Collection}

After carefully reading all 145 papers in full, we extracted the required data and conducted detailed analysis to answer our three research questions. The detailed information is summarized in Table \ref{tab:data}. Data collection mainly concentrated on three aspects: the fundamental information of each paper, some contents about predictive models and the research domain to which each study belongs. In order to prevent data loss and avoid mistakes as much as possible, data collection was performed by the first and fourth authors and then the results were verified by other researchers who are not co-authors of this paper.

\begin{table}[htbp]
\centering\scriptsize
\caption{Data Collection for Research Questions}
\label{tab:data}
 \begin{tabular}{p{0.5cm}p{7.3cm}}
  \toprule
 \textbf{RQs} & \textbf{Types of Data to be Extracted} \\
  \midrule
 RQ1 & Fundamental information for each paper (publication year, author name, type of paper and affiliation).  \\
 RQ2 &  Description, analysis and performance of different predictive models.\\
 RQ3 &  Background, motivation, application scenario, and research topic of each study.\\
  \bottomrule
 \end{tabular}
\end{table}

%% file: RQ1.tex
\section{RQ1: What are the trends in the primary studies on use of predictive models in SE?}

To discuss the emerging trends in predictive model use in SE, in this section we present an analysis of primary studies from three perspectives according to the fundamental information presented in these studies.

\subsection{Publication trends in research on predictive models in Software Engineering}

\begin{figure}[htbp]
\centering\scriptsize
\subfigure[Number of publications per year.]{
\centering
\begin{minipage}[t]{0.47\textwidth}
\includegraphics[width=1\textwidth]{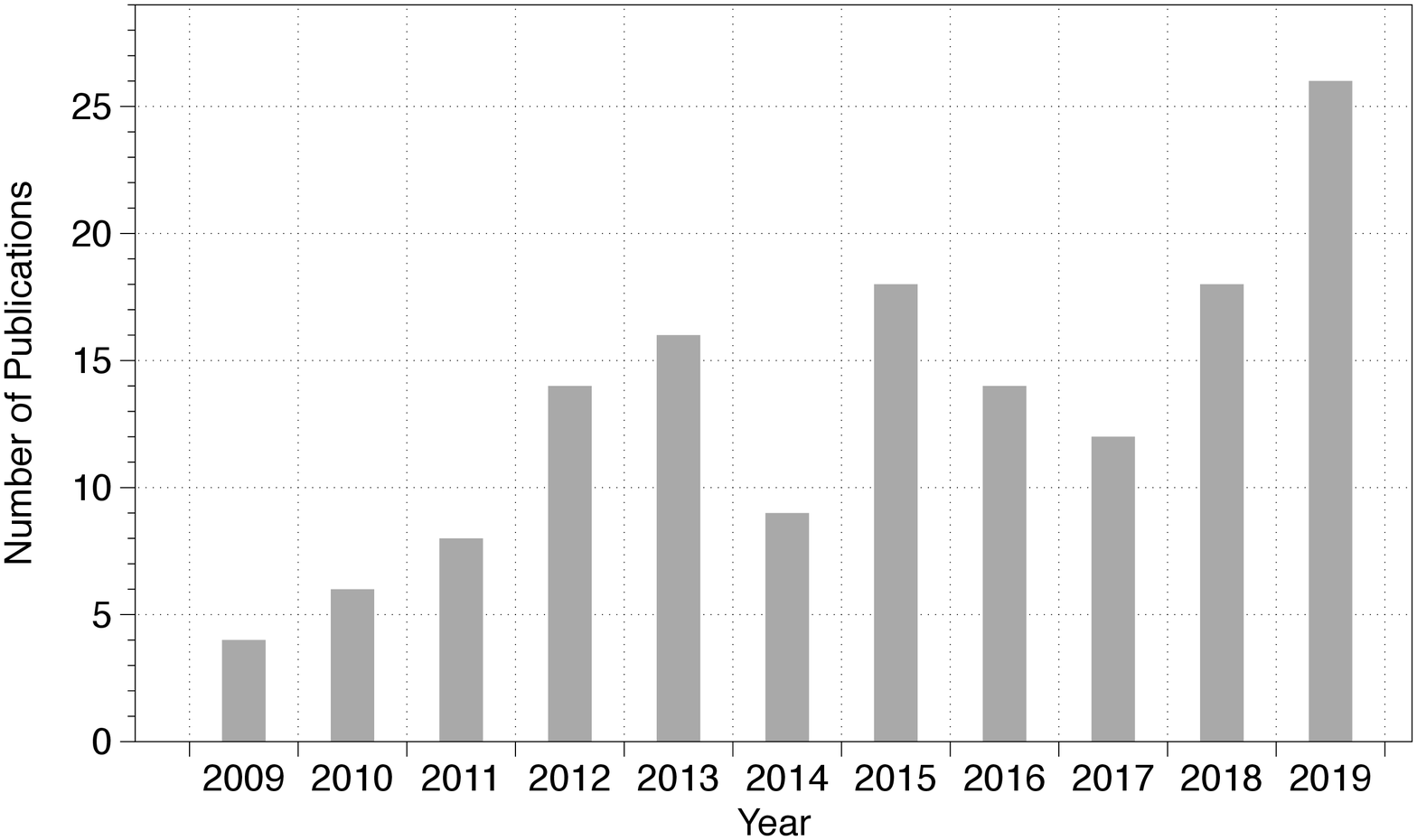}
\end{minipage}
}
%\hfill
\subfigure[Cumulative number of publications per year.]{
%\centering
\begin{minipage}[t]{0.47\textwidth}
\includegraphics[width=1\textwidth]{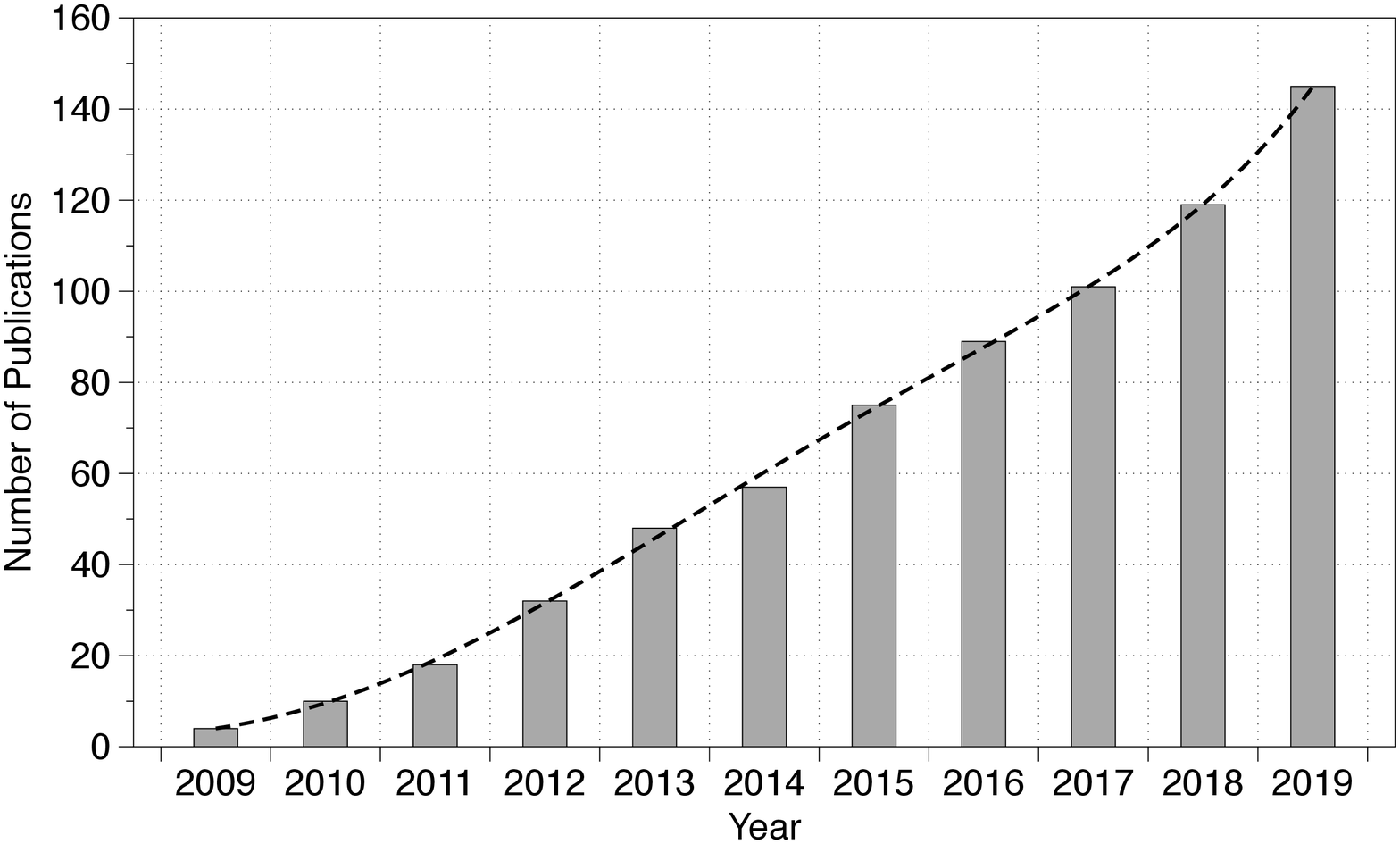}
\end{minipage}
}
\caption{Publication trends of predictive models in the last decade.}
\label{fig.publication trends}
\end{figure}

 Fig. \ref{fig.publication trends}(a) shows the number of papers published between January 1st, 2009 and December 31st, 2019. It is clear that the number of publications from 2009 to 2013 shows a marked increase, with the number reaching 16 publications in 2013. Furthermore, except for 2014 and 2017 with a relatively small number of publications, the number of publications has stayed at a high level and reaches 26 in 2019.

An analysis of the cumulative publications is shown in Fig. \ref{fig.publication trends}(b). We used a polynomial function to fit the cumulative number of publications, revealing the publication trend in the last decade. It can be observed that The ($R^2$) amounts to 0.99803 and the slope of the curve increases markedly between 2009 and 2019. This indicates that research studies using predictive models are likely to continue to experience a strong growth in the future. According to the trend of this curve, it can be foreseen that the cumulative number of publications may be over 170 by the end of 2020. We can see from Fig. \ref{fig.publication trends} that the use of predictive models is becoming more popular in software engineering and there will be more studies that use these models to tackle practical SE problems.

\subsection{Distribution of Publication Venues and Types of Contributions}
The 145 reviewed studies were published in various publication venues, including in 3 top journals and 3 top conferences, well-known and highly regarded in the field of software engineering. The prevalence of papers on predictive models in these conferences and journals indicates construction of predictive models for software engineering purposes are considered of importance. Table \ref{tab:venue} lists the number of papers published in each publication venue. ICSE (47) includes the highest number of primary study papers compared with other top conferences and journals; almost two times higher than ASE (22). %The underlying cause of this phenomenon is the differences in themes between publication venues. JG - I can think of other reasons too... suggest don't need to comment on this?
In Table \ref{tab:venue}, TSE is the most popular journal containing almost 19\% of the primary studies, followed by EMSE and FSE (21). There are 7 relevant studies published in TOSEM. We can  observe that the number of conference papers employing predictive models is higher than that of journal papers.

\begin{table*}[htbp]
\centering\scriptsize
\caption{Top Venues with a Minimum of Two  Predictive Model Based Papers}
\label{tab:venue}
 \begin{tabular}{llp{10cm}c}
  \toprule
 \textbf{Rank} & \textbf{Acronym} & \textbf{Full name} & \textbf{\#Studies}\\
  \midrule
 1. & ICSE & ACM/IEEE International Conference on Software Engineering        & 47  \\
 2. & TSE  & IEEE Transactions on Software Engineering & 27  \\
 3. & ASE  & IEEE/ACM International Conference Automated Software Engineering & 22 \\
 4. & EMSE & Empirical Software Engineering (published by Springer) & 21 \\
 5. & FSE  & ACM SIGSOFT Symposium on the Foundation of Software Engineering/ European Software Engineering Conference & 21 \\
 6. & TOSEM& ACM Transactions on Software Engineering and Methodology & 7 \\

  \bottomrule
 \end{tabular}
\end{table*}

\begin{figure}[htbp]
\centering
%\subfigure[Overall venue distribution.] {
%\centering
%\begin{minipage}[t]{0.4\textwidth}
%\includegraphics[width=1\textwidth]{Figures_Predictive_Models//H.png}
%\end{minipage}
%}
%\hine
%\subfigure[Venue distribution per year.] {
\begin{minipage}[t]{0.48\textwidth}
\includegraphics[width=1\textwidth]{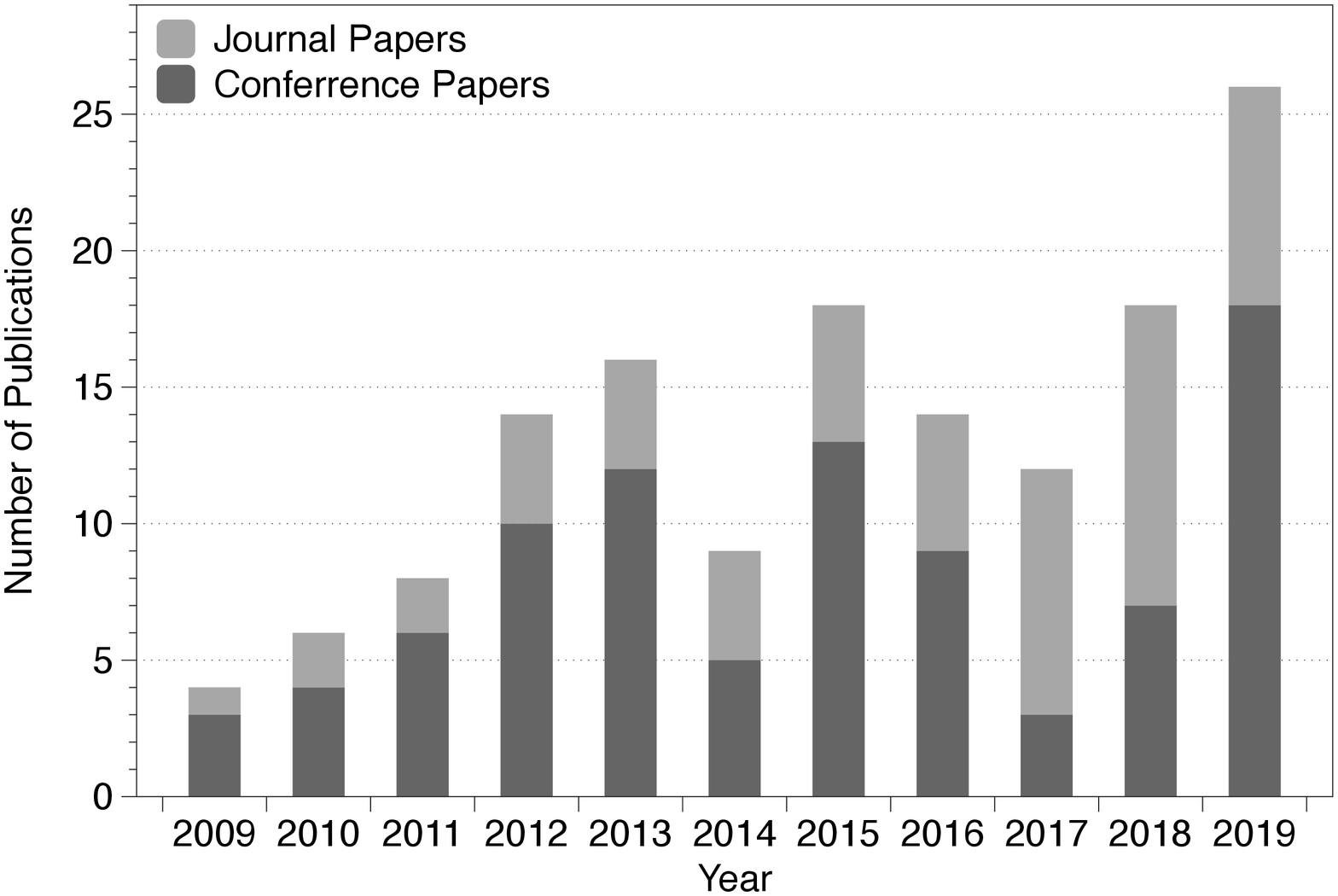}
\caption{Venue distribution per year.}\label{fig:venue}
\end{minipage} 
%\subfigure[Type of main contribution.] {
\begin{minipage}[t]{0.48\textwidth}
\includegraphics[width=.7\textwidth]{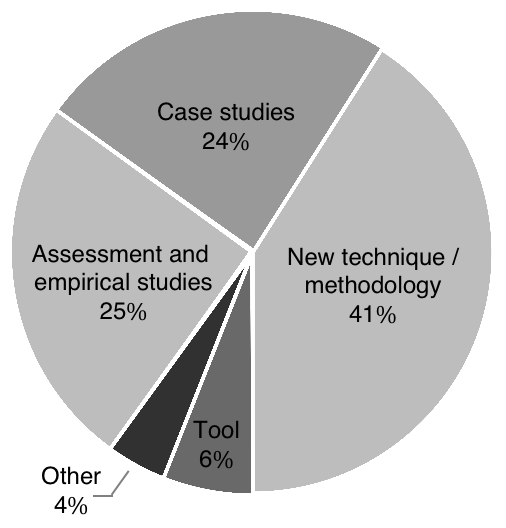}
\caption{Type of main contribution.}\label{fig:type}
\end{minipage}
%}
%\caption{Venue distribution and type of main contribution.}

\end{figure}

We checked the distribution between different publication venues. We find that 62\% of publications appeared in conferences while 38\% were published in journals. Fig. \ref{fig:venue}  gives the venue distribution per year. As From 2009 to 2016, most of the primary studies were published in conference proceedings. This trend has however changed, with the number of journal papers now increasing between 2014 and 2018, often outnumbering conference papers.

%
%\begin{figure}[htbp]
%\centering
%\begin{minipage}[t]{0.49\textwidth}
%\centering
%\includegraphics[width=0.8\textwidth]{Figures_Predictive_Models//G.png}
%\caption{Research domains.}
%\label{fig:domains}
%\end{minipage}
%\begin{minipage}[t]{0.49\textwidth}
%\centering
%\includegraphics[width=1\textwidth]{Figures_Predictive_Models//F.eps}
%\caption{Development trend of five main research domains.}
%\label{fig:development_trend}
%\end{minipage}
%\end{figure}

%\begin{figure}[htbp]
%\centering
%\centering
%\includegraphics[width=0.4\textwidth]{Figures_Predictive_Models//J.png}
%\caption{Type of main contribution.}
%\label{fig.contribution}
%\end{figure}

%\subsection{Types of Contributions}

Fig. \ref{fig:type}  categorizes the primary publications according to their main contribution. The main contribution of 41\% publications was to propose a novel predictive model technique / methodology. 25\% focused on assessment and empirical studies, and 24\% were case studies. The main contribution of 9 studies was to propose a prototype tool (6\%), and 4\% of primary studies were survey or literature reviews on predictive models.

\begin{framed}
\textit{
\textbf{Summary of answers to RQ1:}
 \begin{samepage}
\begin{enumerate}
    \item 145 relevant papers were identified from six top publication venues. The predictive model has attracted sustained interest, with the topic showing steady increase in primary studies in 2009-2019.
	\item Most studies were published at conferences, compared with journals, though the trend changes in recent years. And the main contribution of most studies was to present a new technique or methodology.
 \end{enumerate}
 \end{samepage}
 }
\end{framed} 

%% file: RQ2.tex
\section{RQ2: Which predictive models are applied to software engineering tasks?}

We present an overview of predictive models, including a  classification and research distribution of these models from the primary studies. We also identify several key evaluation metrics that are used in these primary studies for evaluating the performance of their predictive models.

\subsection{Predictive Model Classification}

The predictive model, as a widely-used technology, has many different implementations. Those modelling techniques can be classified into four categories based on \cite{hosseini2017systematic}: (1) Base learners, e.g., naive Bayes, Logistic Regression, (2) Ensemble learning techniques, e.g., Boosting, Bagging, and Random Forest, (3) Neural network, e.g., MLP, Deep Belief Network, CNN, RNN, and LSTM.

\textbf{\textit{Base learners:}}
Base learners are usually generated from training data by a base learning algorithm \cite{zhou2009ensemble}. These are commonly a decision tree, neural network (e.g., MLP, etc.) or other kind of machine learning algorithms. 84 out of 150 primary studies only used base learners to build predictive models. Most of them are classic algorithms which work well for many tasks.

\textbf{\textit{Ensemble learning techniques:}}
Most ensemble methods use a single base learning algorithm to produce homogeneous base learners. However there are also some methods which use multiple learning algorithms to produce heterogeneous learners for reducing bias (Boosting), variance (Bagging), or improving predictions.
The well-known Random Forest algorithm, as a parallel ensemble algorithm, has become one of the most common used ensemble algorithms. A large number of studies (22) employed Random Forest to address different software engineering research tasks. These include defect prediction, vulnerability prediction, code quality prediction, software license exception detection, and fault localization \cite{peters2012privacy, theisen2015approximating, muller2016using, vendome2017machine, zhou2019latent}.

There are 9 studies using a Boosting approach in their reported models (\cite{DBLP:conf/icse/MachalicaSP019, maddila2019predicting, yang2010automatic, falessi2017estimating, malhotra2017empirical, niu2017learning, perini2012machine}). Most of these studies (\cite{yang2010automatic, falessi2017estimating, malhotra2017empirical}) utilized LogitBoost as a part of their approach. Yang et al. \cite{yang2010automatic} implemented a prototype tool with LogitBoost to detecting nocuous coordination ambiguities in requirement documents, which pose a high risk of misunderstanding among different developerss. Falessiet al. \cite{falessi2017estimating} introduced a novel approach to estimate the number of remaining positive links in traceability recovery with LogitBoost. There are some studies that use other Boosting techniques, such as RankBoost and Gradient Boost. Perini et al. \cite{perini2012machine} employed RankBoost in order to give priority to requirements that were to be considered first. Machalica et al. \cite{DBLP:conf/icse/MachalicaSP019} present a data-driven test selection strategy by training a gradient boosted decision tree classifier.

Bagging techniques were also applied by 7 studies to make predictions \cite{zhou2014depth, kim2011dealing}. Zhou et al. \cite{zhou2014depth} leveraged six representative machine learning methods including bagging, AdaBoost and Random Forest to build fault prediction models to study the potentially confounding effect of class size. Kim et al. \cite{kim2011dealing} evaluated the impact of data noise on defect prediction with adoption of basic learners and bagging learners. Malhotra and Khanna \cite{malhotra2017empirical} evaluated the performance of LogitBoost, AdaBoost, Bagging, and other ML techniques for handling imbalanced datasets in software change prediction task.

\begin{table*}[!htbp]
\centering 	
\footnotesize
\caption{The number of predictive models applied in per year.}
\label{tab:predictive models}
 \resizebox{\textwidth}{9.5cm}{
\begin{tabular} {p{2cm} p{1.5cm} p{3.5cm} l l l l l l l l l l l l }
%\hline
\toprule
Category      &Family  & Model Name                         & 2009 & 2010 & 2011 & 2012 & 2013 & 2014 & 2015 & 2016 & 2017 & 2018 & 2019  & Total \\ \midrule%\hline
\multirow{15}{*}{Base learners}
& Regression  & Logistic Regression                & 2    & 3    & 1    & 2    & 6    & 3    & 3    & 3    & 3    & 3    & 1    &30 \\ %\cline{2-13}
& & Multiple linear regression     &     &     &1     &     &     &     & 1    &     &     &     &    &2 \\ %\cline{2-13}
& & Multinomial logistic regression     &     &     &     &     &     &     &     &     &     &2     &    &2 \\ %\cline{2-13}
& & Ordinary Least Squares    &     &     &     &     &     &     & 1     &     &     &     & 1   &2 \\ %\cline{2-13}
& & Binary Logistic Regression     &     &     &     &     &     &     & 1    &     &     &     &    &1 \\ %\cline{2-13}
& & Bayesian Logistic Regression     &     &     &     &     &     &     & 1    &     &     &     &    &1 \\ %\cline{2-13}
& & Bayesian Ridge Regression     &     &     &     &     &     &     &     &     &     &     & 1   &1 \\ %\cline{2-13}
& Naive Bayes & Naive Bayes                        &     & 3    & 3    & 4    & 4    & 3    & 4    & 3    & 2    & 3    & 1   &30  \\ %\cline{2-13}
&  & Bayesian Network                  &     & 1    & 2    &     &     & 1    & 1    &     & 1    & 2    & 1    &9 \\ %\cline{2-13}
& & Multinomial Naive Bayes                        &     &     &     &     &     &     &     &     &     & 1    &    &1  \\ %\cline{2-13}
&SVM & SVM                                &     & 2    & 3    & 3    & 5    & 2    & 2    & 2    & 1    & 3    & 4    &27 \\ %\cline{2-13}
& & Linear SVM                                &     &    &     &     &     &     & 1    &     &     &     & 1    &2 \\ %\cline{2-13}
& & SMO                               &     &    &     &     &     &     & 1    & 1     &     &     &     &2 \\ %\cline{2-13}
& Decision Tree & Decision Tree                      &     & 2    & 1    & 2    & 1    & 1    &     & 1    &     & 1   & 4    &13 \\ %\cline{2-13}
& & J48                     &     & 1    &     &     & 1    &     & 4    & 2    & 1    & 1    &     &10 \\ %\cline{2-13}
& & C4.5                     &     &     &1     &     &     & 1     & 2    &     &     & 1    &     &5 \\ %\cline{2-13}
& & ADTree                     &     &     &     &     & 1     &      & 1     &      &     & 1     &     &3 \\ %\cline{2-13}
& & CART                     &     &     &     &1     &      &      &      &      &3     &      &     &2 \\ %\cline{2-13}
&  & Alternating Decision Tree  &      &      &      &      & 1    &      &      &      &      &      &      &1 \\ %\cline{2-13}
&  & Recursive Partitioning     &      &      &      &      &      &      &      & 1    &      &      &      &1 \\ %\cline{2-13}
& Clustering technique  & K-means                            &     &     &     & 1    & 2    &     & 1    & 1    & 1    &     & 1   &7  \\ %\cline{2-13}
&  & Expectation Maximization               &      &      &1      &      &      &      &1      &     &      &      &     &2  \\ %\cline{2-13}
&  & Spectral Clustering                &      &      &      &      &      &      &      & 1    &      &      &     &1  \\ %\cline{2-13}
&  & Fuzzy C- means                     &      &      &      &      &      &      &      & 1    &      &      &     &1  \\ %\cline{2-13}
&  & Partitioning Around Medoids                    &      &      &      &      &      &      &      & 1    &      &      &     &1  \\ %\cline{2-13}
& Nearest Neighbor & K-nearest Neighbor                 &     &     &     &     & 1    & 1    &     & 1    & 1    & 1    & 1    &6 \\ %\cline{2-13}
&  & Nearest Neighbor Classifier        &     &     &     & 1    & 1    &     & 1    & 1    & 1    &     &     &5 \\ %\cline{2-13}
&  & nearest neighbor with non-nested generalization (NNGe)       &     &     &     & 1    &     &     &     &     &     &     &     &1 \\ %\cline{2-13}
& Discriminative Model & Markov Chains                      &     &      & 1    &      &     &      &1      &      &      &      &1      &3 \\
&  & Conditional Random Field           &      &      &      &      &      &      &      &      &      & 2    &     &2 \\ %\cline{2-13}
& Rule-based & Conjunctive Rule                    &      &      &      &      &      & 1      & 1     &     &      &      &     & 2  \\ %\cline{2-13}
& & Ripper         &      &      &      &      &      &       & 1     & 1    &      &      &     & 2  \\ %\cline{2-13}
& & Ridor         &      &      &      &      &      &       & 1     &     &      &      &     & 1  \\ %\cline{2-13}

%&  & Tree Bagging             &      &      &      &      &      &      &      & 1    &      &      &      &1 \\ %\cline{2-13}
\midrule%\hline
\multirow{5}{*}{Ensemble learning}
& Random Forest  & Random Forest                      &      &      &      &      &      & 1    & 3    & 6    & 4    & 6    & 3    & 23 \\ %
& Boosting & Tree Boosting                           &      &      &      &      &      &      &      &      &      & 2    &1     &3  \\ %\cline{2-13}
& & AdaBoost                           &      &      &      &      &      & 1    &      &      &      & 1    &     &2 \\ %\cline{2-13}
 & & XGBoost                            &      &      &      &      &      &      &      &      &      &1      & 1    &2 \\ %\cline{2-13}
  & & LogitBoost                            &      & 1      &      &      &      &      &      &      &      &      &     &1 \\ %\cline{2-13}
    & & gradient boosting                          &      &       &      &      &      &      &      &      &      &      & 1     &1 \\ %\cline{2-13}
 & Bagging & Bagging                            &      &      & 1    & 1      &      &      & 1     & 2     &      & 2    &     &7 \\ %\cline{2-13}
 \midrule%\hline

 \multirow{8}{*}{Neural network}
 & Perceptron model &  Multi-Layer Perceptron (MLP)                              &      &      &      &      & 1     &      & 1     &      & 1     & 1    & 2    &6 \\ %
 &  & Single-Layer Perception          &      &      &      &      &      &      &      &      &      & 1    &      &1 \\ %
& MP Neuron model & ANN                              &      &      &      &      &      &      & 1     &      & 1     &     & 2    &3 \\
&  & AVNNet          &      &      &      &      &      &      &      & 1     &      & 1    &      &2\\
&  & Siamese Neural Network (SNN)          &      &      &      &      &      &      &      &      &      & 1    &1      &2\\ %  %

& Deep learning techniques & LSTM                               &      &      &      &      &      &      &      &      &      & 1    & 4    &5 \\ %\cline{2-13}
& & RNN                                &      &      &      &      &      &      &      &      &      & 1    & 1    &2 \\
&  & Bi-LSTM                            &      &      &      &      &      &      &      &      &      &      & 1    &1 \\ %
&  & GRU              &      &      &      &      &      &      &      &      &      &      & 1    &1 \\ %\cline{2-13}
&  & Recurrent Highway Network          &      &      &      &      &      &      &      &      &      & 1    &      &1 \\ %
&  & CNN                                &      &      &      &      &      &      &      & 1    &      & 2    & 1    &4 \\ %\cline{2-13}
&  & Deep Belief Network (DBN)          &      &      &      &      &      &      &      & 1    &      &      &      &1 \\ %\cline{2-13}
%\cline{2-13} \cline{2-13}
\bottomrule
\end{tabular}}
\end{table*}

%\hl{\textbf{\textit{Neural network:}}
%Neural networks (NN) are a set of algorithms (e.g., perceptions) that are designed to identify patterns, which can be classified into two categories: Single-Layer Neural Network and Deep Neural Network. They interpret sensory data through a machine perception, labeling or classifying raw input. Therefore, NN can be considered as components of larger machine-learning applications involving algorithms for clustering, classification or regression. There are over 10 studies applying different NN techniques to tackle different SE problems.}

\textbf{\textit{Neural Network:}}
Neural networks (NN) are a set of algorithms (e.g., perceptions and deep learning techniques) designed to identify patterns. They interpret sensory data through a machine perception, labeling, or classifying raw data input. Therefore, NN can be considered as components of larger machine-learning applications involving algorithms for clustering, classification, or regression. Neural Network models can be classified into three categories, i.e., perceptron models, MP Neural models, and deep learning models, through the analysis of selected studies.

A perceptron is the simplified form of a neural network applied for supervised learning of binary classifiers, which consists of four main components including input values, net sum, weights and bias, and an activation function.

The MP neuron model is actually a simplified model constructed based on the structure and working principle of biological neurons, which consists of a function with a single parameter, taking binary input, and giving binary output according to a determined threshold value.

Deep learning is part of a broader family of machine learning methods based on artificial neural networks \cite{schmidhuber2015deep}. Deep learning architectures, such as deep neural networks, deep belief networks, recurrent neural networks (RNN), and convolutional neural networks (CNN), have been applied to help various tasks across the software development life cycle. With the advent and development of deep learning techniques, 15 primary studies utilized deep learning techniques in their reported models \cite{lin2018sentiment, cambronero2019deep, liu2018deep, xu2016predicting}. In this section, we briefly introduce several deep neural networks to give an overview of the fundamental principles. We also present how they have been applied to solve software engineering problems.

CNN has been used for SE tasks, e.g., \cite{liu2018deep}. The CNN model consists of tree layer, including an input layer, a hidden neurons layer and an output layer. CNN has achieved significant advances in recent years, with CNNs effectively increasing the flexibility and capacity of machine learning \cite{liu2018deep}. A CNN predictive model is made up of two parts, i.e., feature extraction and prediction. Compared to other models, the biggest strength of CNN lies in its convolution kernels, which focus only on local features and can fully extract the internal features of the data, boosting its accuracy. CNN have been applied to many SE research tasks. Liu et al. \cite{liu2018deep} exploited CNN to identify feature envy smells. The model they build had three layers with 128 kernels. The performance  of their proposed approach outperformed the state-of-the-art significantly in feature envy detection and recommendation. Xu et al. \cite{xu2016predicting} present a deep-learning based approach to predict semantically linkable units in developers’ discussions in Stack Overflow. They formulated this problem as a multiclass classification problem, training a CNN to solve it. For classifying semantic relatedness, they used filters of five different window sizes, and each window size contained 128 filters to capture the most informative features.

Another approach is to use an LSTM model, a variant of the Recurrent Neural Network (RNN). This is specially designed for precessing sequential data. An LSTM model can capture the contextual information of sequences thanks to the recurrent nature of the RNN \cite{zhang2019robust}. Computational units connected in each layer form a directed graph along a temporal sequence, which allows it to exhibit temporal dynamic behavior. Each LSTM unit contains an input gate, a memory cell, a forget gate, and an output gate. The gating mechanism of LSTM guarantees that the gradient of the long-term dependencies will not vanish \cite{chen2019sentimoji}.

Zhang et al. \cite{zhang2019robust} utilized an attention-based Bi-LSTM model to detect anomalies in log events. Different from a standard LSTM, a Bi-LSTM can divide the hidden neuron layer into forward and backward, which allows it to capture more information of the log sequences. They confirmed the effectiveness of the proposed approach by comparing it with several traditional machine learning methods. Chen et al. \cite{chen2019sentimoji} leveraged Bi-LSTM model as an emoji-powered learning approach for sentiment analysis. In their paper, the Bi-LSTM model contained two bi-directional LSTM layers and one attention layer, which treated the sentence vectors in Twitter and GitHub posts containing emojis as inputs and output the probabilities that instances contain each emoji.

\subsection{Distribution of Different Predictive Models}

In order to get a better understanding of  predictive models, we summarize the  frequency of models used in different software engineering tasks. Table \ref{tab:predictive models} shows the number of cases for which different categories of predictive models and relevant studies in which they appear. Comparing with another two categories, the base learner is the most commonly used predictive models among all learners including eight different families, where there are 39 and 38 studies employing Logistic Regression and Naive Bayes to address software engineering problems in the last decade. SVM and Decision Tree are the next most frequently used predictive models with 31 and 35 studies using them, followed by Nearest Neighbor. Discriminative models and rule-based  learners are also applied to software engineering tasks with high probability, including Markov Chains, CRF, Conjunctive Rule-based model, Ripper, and Ridor. It can be seen from Table \ref{tab:predictive models} that the usage of base learners has remained stable between 2009 and 2019, indicating that these learners are still effective techniques in solving suitable problems, although they have been introduced some time ago.

There are 39 studies applying ensemble algorithms that can be classified into three categories, involving Boosting, Bagging, and Random Forest. 59\% of the studies selected Random Forest as the predictive model. 9 studies leveraged Boosting techniques in their experiments, in which 3 used TreeBoosting and 2 studies each used AdaBoost and XGBoost. Another 2 studies adopted LogisticBoost and gradient boosting respectively to address specific problems. Bagging is another popular ensemble technique being used in 7 software engineering primary studies.

%need a paragraph to express the trend of neural network

%\hl{Neural Networks ,as one of the most popular techniques in recent years, are applied in 14 primary studies. Perception models are the most commonly used with 7 studies using them, where 6 studies selected to apply Multi-Layer perception model as their predictive model. There are 3 studies adopted Artificial Neural Network (ANN) while Siamese Neural Networks (SNN) and AVNNet were used twice respectively among all studies.}

There are 29 studies that adopted Neural Network techniques, where 22 of the 29 studies used deep learning techniques, and most of these studies were conducted in 2018 and 2019. This suggests the application of deep learning to software engineering is exhibiting a booming trend in the last few years. In Table \ref{tab:predictive models}, the family of deep learning techniques, including LSTM, RNN, Bi-LSTM, GRU, and Recurrent Highway Network, is very popular with 10 studies applying them to software engineering problems. The second commonly used deep learning model is the CNN, with 4 studies using it. Deep Relief Network (DBN) was used only once among all selected studies. 7 studies adopted perceptron models, where Multi-Layer Perceptron was used in 6 studies. There are also 7 studies that employed MP neuron models, in which 3 used ANN, 2 used AVNNet, and another 2 studies used SNN.

%CNN is the third most commonly used deep learning technique with 4 studies using it.

\subsection{Evaluation Metrics for Predictive Models}

 Table \ref{tab:measures} lists the most widely used metrics in the primary studies, including a description, definition and the related studies. As Table \ref{tab:measures} shows, Recall, Precision and F-measure are the most commonly used to evaluate the performance of predictive models, followed by Accuracy and AUC. 9 of the studies have used the probability of false alarm (pf) and 6 studies have used balance. ROC and MAE were applied in 5 studies each. Finally, only 4 studies selected G-measure and 3 selected specificity as evaluation measures.

 \begin{table*}[!htbp]
\centering
\scriptsize
\caption{The definition and usage of common evaluation measures across the primary studies.}
\label{tab:measures}
\begin{tabular}{ p{3cm} p{7cm} l c }
%\hline
\toprule
\textbf{Measure} & \textbf{Description} & \textbf{Definition} & \textbf{\#Studies}  \\ \midrule%\hline
Recall & The proportion of positive cases that were correctly predicted & $\frac{TP}{TP + FN}$ & 96
\\ %\cline{2-13}
Precision & The proportion of predicted positive instances that are correct. & $\frac{TP}{TP+FP}$ & 89
\\ %\cline{2-13}
F-measure & Harmonic mean of precision and recall & $\frac{2 \times Precision \times Recall}{Precision + Recall}$ & 68
\\
\\
Accuracy & Proportion of correctly classified instances & $\frac{TP + TN}{TP + FP + TN + FN}$ & 32
\\
Area Under the Curve (AUC) & The area under the receiver operating characteristics curve. Independent of the cutoff value & & 30
\\
Probability of False Alarm (pf) & Proportion of negative instances incorrectly classified as positive & $\frac{FP}{FP + TN}$ & 9
\\
Balance(bla) &  The balance between pd (the probability of detection) and pf. The bigger the bal value is, the better the performance of learning model is. & $1 - \frac{\sqrt{(1 - pd)^2 + (0 - pd)^2}}{\sqrt{2}}$ & 6
\\
ROC & A comprehensive indicators reflecting sensitivity and specificity of continuous variables & & 5
\\
Mean Absolute Error (MAE) & The average of absolute errors, which can better reflect the actual situation of the predicted value error. &$\frac{1}{m}\sum_{i=1}^{m}{(y_i - \widehat{y})^2}$ & 5
\\
G-measure & Harmonic mean of pd and (1-pf) & $\frac{2 \times pd \times (1 - pf)}{pd + (1 - pf)}$ & 4
\\
Specificity & The proportion of predicted negative instances that are correct & $\frac{TN}{FP + TN}$ & 3
\\
\bottomrule
\end{tabular}
\end{table*}

 \begin{framed}
\textit{
\textbf{Summary of answers to RQ2:}
 \begin{samepage}
\begin{enumerate}
	%\item Predictive models can be classified into four categories, i.e., base learners, ensemble learning techniques, Neural Network and deep learning techniques.
	\item Most of the primary studies applied base learners to tackle software engineering problems, where Logistic Regression and Naive Bayes are the most popular predictive models. Using deep learners in software engineering shows a thriving trend in recent years.
	\item Recall, precision and F-measure are the three most commonly used evaluation metrics to evaluate the performance of different predictive models.
 \end{enumerate}
 \end{samepage}
 }
\end{framed} 

%% file: RQ3.tex
\section{RQ3: In what domains and applications have predictive models been applied?}

Of the 145 studies in our survey, over 80\% involved applying approaches using predictive models to a specific software engineering problems. In this section, we conduct an analysis on the distribution of different research domains to which predictive models were applied, and analyse the application of predictive models in each research topics.

To better understand the application distribution of predictive models, we categorise the software development life cycle (SDLC) into the following stages according to \cite{bourque2014guide}: (1) Software Requirements; (2) Software Design and Implementation; (3) Software Testing and Debugging; (4) Software Maintenance; and (5) Software Engineering Professional Practice.

We referred to the definition of each category of the SDLC in \cite{bourque2014guide} and conducted a comprehensive analysis of the problem addressed by each study and its practical application scenarios to determine which category the study belongs to.

\subsection{Distribution in Different SDLC Domains}

\begin{figure}[htbp]
\centering
\begin{minipage}[t]{0.49\textwidth}
\centering
\includegraphics[width=0.8\textwidth]{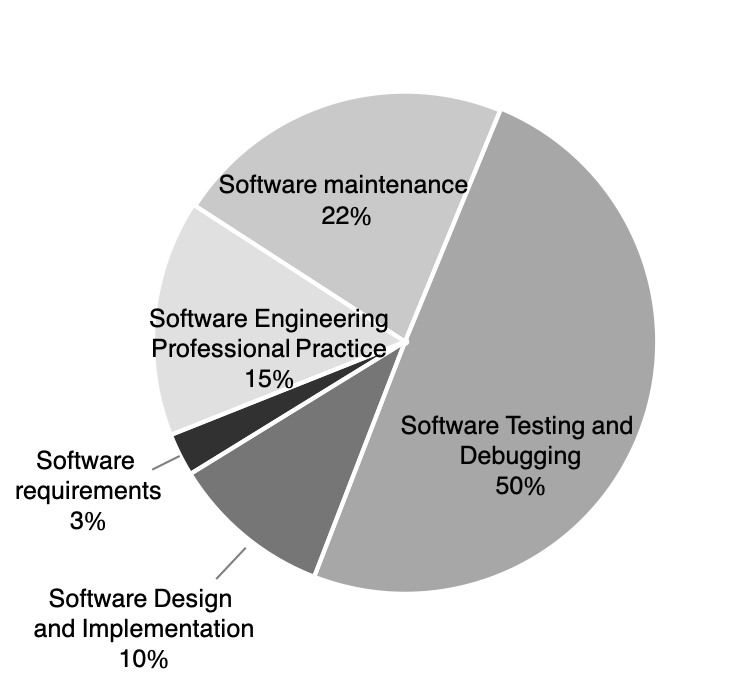}
\caption{Research domains.}
\label{fig:domains}
\end{minipage}
\begin{minipage}[t]{0.49\textwidth}
\centering
\includegraphics[width=1\textwidth]{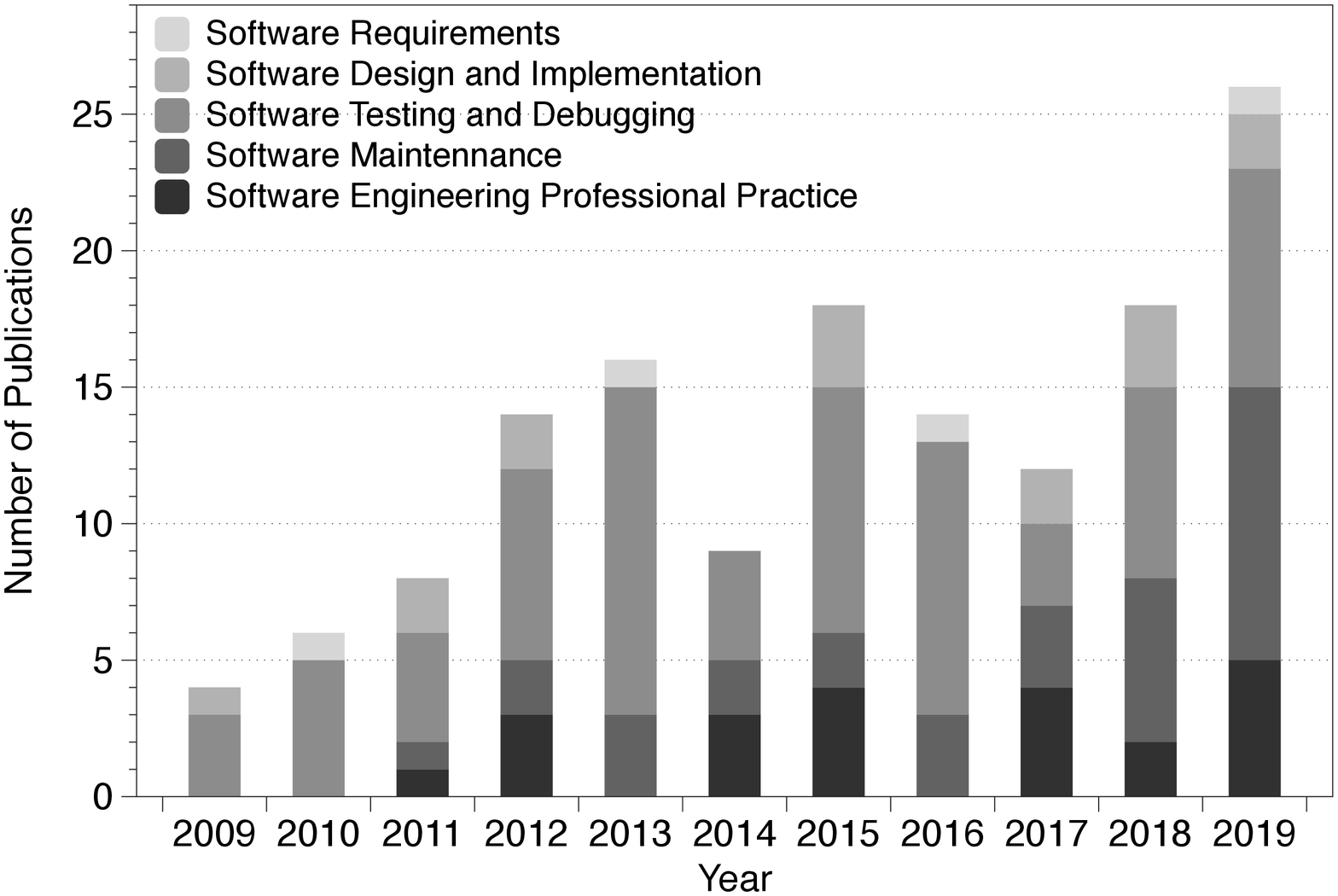}
\caption{Development trend of five main research domains.}
\label{fig:development_trend}
\end{minipage}
\end{figure}

We categorize our selected primary studies based on research domains in Fig. \ref{fig:domains} and illustrate the development trend of the main research domains in each year in Fig. \ref{fig:development_trend}. Table \ref{tab:topics} lists the ranking of software engineering tasks that most studies have concentrated on.

As Fig. \ref{fig:domains} shows, the primary research domain of most of the studies (50\%) was software testing and debugging, followed by software maintenance (22\%). Studies related to software engineering professional practice were 15\%. Software design and implementation were the focus of 10\% of the studies. Finally, only 8\% of the papers focus on software requirements (3\%). We notice that most of the primary studies worked on specific tasks with respect to software testing and maintenance. One potential reason is that these two research directions involve many predictive model-applicable tasks, such as defect prediction, code smell detection, and bug report management.

\begin{table*}[!htbp]
\centering\scriptsize
\caption{Top Software Engineering Topics with a Minimum of Five Predictive Model Based Papers}
\label{tab:topics}
\footnotesize
\begin{tabular}{ c p{6cm} p{5.5cm} c }
%\hline
\toprule
 \textbf{Rank} & \textbf{Specific Task} & \textbf{Research Domain}  &  \textbf{\#Studies} \\
\midrule%\hline
1. & Defect Prediction & Software Testing and Debugging & 40
\\
2. & Bug Report Management & Software Testing and Debugging & 13
\\
2. & Software Quality Management & Software Testing and Debugging & 13
\\
4. & Software Effort/Cost Prediction  & Software Design and Implementation & 10
\\
5. & Developer Behavior Analysis & Software Engineering Professional Practice & 9
\\
6. & Code Clone Management  & Software Maintenance & 6
\\
7. & Software Change Management & Software Maintenance & 6
\\

\bottomrule
\end{tabular}
\end{table*}

Fig.\ref{fig:development_trend} shows the development trend of the main research domains -- Software requirements, Software development, Testing, Maintenance, quality assessment and Software engineering professional practice -- over the last decade. Apart from 2009, there are a few studies related to software engineering professional practice in each year. Only 4 software requirement related studies using predictive models appeared in last decade. This indicates that software engineering professional practice shows a stable development trend, but there seems to be a lack of attention to predictive model research topics in software requirements. The proportion of studies in software maintenance shows an increase in recent years. We notice from Fig.\ref{fig:domains} that predictive models have not yet been widely used in certain domains, such as software requirements and software design. This may perhaps motivate researchers to explore more scenarios in these SDLC domains suitable for application of predictive models.

Table \ref{tab:topics} lists the ranking of specific topics where at least five predictive model related papers have appeared. We observe that five of the seven research topics belong to software testing and debugging (3) as well as software maintenance (2), which is consistent with the proportion of this research domain in Fig.\ref{fig:domains} and Fig.\ref{fig:development_trend}. There are 40 publications employing predictive models in defect prediction, a research topic that most commonly to use predictive models as a solution, followed by but report management (13) and software quality management (13). The main contribution of 10 primary studies was to perform software effort / cost prediction by building predictive models. The number of publications working on code clone management is the same as that of software change management. Besides, 9 studies concentrated on developer behavior analysis (9).

\subsection{Software Requirements}

Requirements engineering (RE) is the process of defining, documenting, and maintaining requirements in software design process \cite{bourque2014guide} and play a crucial role throughout systems' lifecycle. A requirement with high quality can notably improve development efficiency by avoiding reconstruction and rework. %With the constant specification of the software development process, a growing number of studies concentrated on requirement-related tasks. - the data doesn't show this! JG

\subsubsection{Requirements classification}
 An effective classification of software requirements enables analysts to perform well-focused communication with developers and users as well as prioritize the requirement documents according to their importance \cite{abad2019supporting}.

\textbf{Non-functional requirements (NFRs) classification:} Anish et al. \cite{anish2016probing} conducted a study to identify, document, and organize Probing Questions (PQs) for five different areas of functionality into structured flows, called PQ-flow. They used Naive Bayes to identify Architecturally Significant Functional Requirements (ASFRs), used random k labelsets classifier (RAkEL) to categorize ASFR by types, and finally recommended PQ-flows.

%We report a new study with over 40 experienced architects to identify reusable PQs for five areas of functionality and organize them into structured flows.  These PQflows can be used by Business Analysts to elicit and specify architecturally relevant information. Additionally, we leverage machine learning techniques to determine when a PQ-flow is appropriate for use in a project, and to annotate individual PQs with relevant information extracted from the existing SRS.

\textbf{Other requirements classification:} Abad et al. \cite{abad2019supporting} proposed a novel approach that extracts and classifies requirements-related knowledge to support analysts familiarity with different domains.  Based on the intuition behind utilizing lexical association, they built a dynamic generative model to extract the relevant terms in documents and applied the rational kernels method as well as SVMs when classifying requirements. %Although their proposed method has been evaluated, some factors (e.g. human interface factors) might impact the performance of this technique in industrial contexts.

\subsubsection{Coordination of Requirements Detection}

In large software projects, participants engage in parallel and interdependent tasks, resulting in work dependence and coordination needs. When developers remain unaware or do not obtain timely awareness of the coordination that is required to manage work dependencies, there is potential for software productivity and quality problems. Yang et al. \cite{yang2010automatic} implemented a prototype tool for Nocuous Ambiguity Identification (NAI) in text. In their earlier work, they had focused on identifying the coordination ambiguity and anaphora ambiguity in requirement documents, and thus this study tried to detect another two types of ambiguity (i.e., nocuous and innocuous ambiguity). After extracting ambiguity instances and recognizing coordination constituents, they defined a set of heuristics (e.g., coordination matching, distribution similarity, and collocation frequency) based on word distribution and collocation and trained a binary classifier using LogitBoost algorithm to classify the input as nocuous or innocuous. Blincoe et al. \cite{blincoe2013all} investigated what work dependencies a software development team needs to consider when establishing the coordination needs. %They present a list of properties helping to characterize task pairs that require coordination. They noticed that existing techniques for detecting coordination requirements find excessive dependencies.
They applied k-nearest neighbor machine learning algorithm to identify and  supplement task properties that are indicative of the crucial coordination needs, contributing in the selection of the most critical coordination needs.

\subsection{Software Design and Implementation}

A software design is a description of the architecture of the software to be implemented, and algorithms or code management used. Below we  introduce the application of predictive models in software design and for code development.

\subsubsection{Software Architecture and Design}

The architecture of a system describes the relationships between its main components and how they interact with each other. There are many factors involved in software architecture and design (e.g., quality, security, performance, etc.). It is a challenging task  to balance different factors for designing a good architecture. To identify a set of quality concerns for maintaining security, reliability and performance of software systems, Mirakhorli et al. \cite{mirakhorli2012tactic} proposed a cost-effective approach to automatically construct traceability links for architectural tactics. They applied a base learner as a tactic-classifier identifying all classes related to a given tactic.  To reconstruct tactic-level traceability, these classes were mapped to their relevant tactic through two different classifiers, where one was  trained using textual descriptions of each tactic and another using code snippets taken. %Their experimental results showed that code-trained classifiers outperformed description-trained classifiers in terms of precision and recall.

Architects construct software architectures by making a variety of design decisions that can satisfy quality concerns (e.g., reliability, performance, and security). However, the architectural quality decreased when developers modified code without fully understanding design decisions. In order to address this problem, Mirakhorli et al. \cite{mirakhorli2015detecting} proposed a solution for detecting, tracing and visualizing architectural tactics in code. They utilized six machine learning algorithms (i.e., SVM, Decision Tree (C.45), Bayesian Logistic Regressions (BLR), AdaBoost, SLIPPER, Bagging) to train classifiers for detecting the presence of architectural tactics in source code. They monitored the architecturally significant code by mapping those relevant code segments into Tactic Traceability Patterns (tTPs) and notified developers when those segments were modified. 
%By analyzing the experimental results, they concluded that the six of classifiers performed equivalently in tactical detection tasks.

Gopalakrishnan et al. \cite{gopalakrishnan2017can} present a bottom-up approach to identify architectural tactics. They adopted the Random Forest algorithm to capture relationships between latent topics and architectural tactics in source code of projects. 
%They evaluated the ability of the proposed approach through a series of experiments on two large-scale datasets, i.e., Apache Hive and Hadoop. 
Reman et al. \cite{rehman2018roles} researched software architecture in another perspective. They performed a case study to investigate whether and how Hands-on software architects benefit the projects by using data analytics-based techniques in five large-scale industrial systems. To better understand what type of code architects write, they employed a classification algorithm \cite{mirakhorli2012tactic}  to identify functional comments and analyzed architects' code contributions. %They observed that architects write code for addressing quality concerns and shaping the initial structure of projects. %They also noticed that some architects engaged in other activities in addition to writing tactic, functional code as well as tests.

\subsubsection{Software Development Effort Estimation}

%In software development, effort estimation is the process of predicting the most realistic amount of effort required to develop or maintain software, which can provide beneficial suggestions for project plans, iteration plans, budgets, investment analyses, pricing processes and bidding rounds.

In software development, Software effort estimation (SEE) is the process of predicting the most realistic amount of effort required to develop or maintain software \cite{jorgensen2014we}. SEE is an important activity in the software development process since it can provide beneficial suggestions for project plans, iteration plans, budgets, investment analyses, pricing processes and bidding rounds \cite{jorgensen2014we}. There are many studies related to effort estimation \cite{nan2009impact, choetkiertikul2015predicting}. To estimate SDLC, Nan et al. \cite{nan2009impact} formalized the nonlinear effect of management pressures on project performance as U-shaped relationships using Regression Models. They found that controlling for software process, size, complexity, and conformance quality, budget pressure, a less researched construct, has significant U-shaped relationships with development cycle time and development effort. Choetkiertikul et al. \cite{choetkiertikul2015predicting} proposed a novel approach to provide automated support for project managers and other decision makers in predicting whether a subset of software tasks in a software project have a risk of being delayed. They used random forest as the local classifier on both local data and networked data. %In addition, they use d collective classification to simultaneously predict the degree of delay for a group of related tasks.

More than half the literature on SEE has concentrated on comparisons of new estimation methods \cite{choetkiertikul2018deep, menzies2017negative, malgonde2019ensemble}. \cite{dejaeger2011data, mittas2012ranking} conducted comparative studies to investigate which estimation approach performs the best for software effort estimation. Whigham et al. \cite{whigham2015baseline} proposed an automatically transformed linear model (ATLM) as a suitable baseline model for comparison against software effort estimation methods. ATLM performed well over a range of different project types. Moreover, ATLM was able to be used with mixed numeric and categorical data and requires no parameter tuning.  Dejaeger et al. \cite{dejaeger2011data} conducted a comparative study of data mining techniques for software effort estimation. Their techniques includes tree/rule-based models, linear models such as various types of linear regression, nonlinear models (MARS, multilayered perceptron neural networks, radial basis function networks, and least squares support vector machines), and estimation techniques that do not explicitly induce a model (e.g., a case-based reasoning approach). Their results showed that least squares regression in combination with a logarithmic transformation performs best. Mittas et al. \cite{mittas2012ranking} proposed a statistical framework based on a multiple comparisons algorithm in order to rank several cost estimation models. They identified those which have significant differences in accuracy and clustered them into non-overlapping groups. They examined the predictive power of 11 models over 6 public domain datasets. Their results showed that very often a linear model is adequate enough, but not the optimal solution. Song et al. \cite{song2018novel} aimed to present a synthetic data generator to alleviate the data scarcity problem of software effort estimation. They compared their approach with six state-of-the-art models to evaluate its performance. Maddila et al. \cite{maddila2019predicting} constructed a SEE model on an individual pull request level, helping developers trace their work items. Their methodology was deployed by several thousand developers across several software families. They observed from their deployment that 42.94\% of developer time was saved by deploying their approach.

\subsubsection{Software Process Evaluation}

Software process evaluation is essential to improve software development and the quality of software products in an organization. Chen et al. \cite{chen2011software} formulated this problem as a sequence classification task and introduced a novel semi-automated approach utilizing three well-known and representative classification algorithms (i.e.,  decision trees (C4.5), Naive Bayes (NB) algorithm, and Support Vector Machines (SVM)). Compared with traditional approaches, it was time-saving and avoided inaccurate evaluation results even if not each artifact can be accessed directly due to privacy or security concerns.

%For example, if an external audit agency or a consultation agency is asked to evaluate the software processes of an organization, it may not be allowed to access every artifact directly due to privacy or security concerns. This may lead to inadequate checking of artifacts, which result in inaccurate evaluation results.

\subsection{Software Testing and Debugging}

%Testing is the process of verifying and validating that a software or application is bug free, meets the technical requirements as guided by its design and development and meets the user requirements effectively and efficiently with handling all the exceptional and boundary cases.

%Debugging is the process of fixing a bug in the software. It can defined as the identifying, analyzing and removing errors. This activity begins after the software fails to execute properly and concludes by solving the problem and successfully testing the software. It is considered to be an extremely complex and tedious task because errors need to be resolved at all stages of debugging.

Software testing is the process of verifying and validating whether a software or application is bug free, and its design and development meet technical and user requirements. Debugging  is the process of identifying, analyzing and remove errors after the software fails to execute properly. Effective testing and debugging techniques can improve software quality with reduced effort consumption and improved performance \cite{tian2018deeptest}.

\subsubsection{Test Report Classification}

Podgurki et al. \cite{podgurski2003automated} present an automated approach for classifying failure reports with closely related causes. Their proposed classification strategy involved supervised and unsupervised pattern classification, as well as multivariate visualization. Wang et al. \cite{wang2016local} introduced a Local-based Active ClassiFication (LOAF) approach to classify crowdsourced test reports that reveal true faults. They adopted active learning to train a classification model with as few labeled inputs as possible in order to reduce the onerous burden of manual labeling. Compared with existing supervised machine learning methods, LOAF exhibited promising results on one of the largest Chinese crowdsourced testing platforms.

\subsubsection{Test Case Quality Management}

In software testing, test case quality is a critical factor. Good test cases can effectively improve software quality. On the contrary, constructing and running bad test cases may cost significant time and effort while still not finding defects or faults. There are many studies that aim to improve test case quality using predictive models.

 Natella et al. \cite{natella2012fault} proposed a new approach to refine the ``fault load" by removing faults that are not representative of residual software faults. They used a decision tree to classify whether a fault is representative or not based on software complexity metrics. Fard et al. \cite{milani2014leveraging} implemented a tool, called Testilizer, to automatically generating test cases for web applications. Since the problem of generating similar test suites was viewed as a classification task, they present a set of features and trained a classifier by leveraging human knowledge mining from human-written test suites.

 Cotroneo et al. \cite{cotroneo2013learning} proposed a method based on machine learning to adaptively combine testing techniques during the testing process. Their method contains an offline learning phase and an online learning phase. During offline learning, they first defined the features of a testing session potentially related to the techniques performance, and then used several machine learning approaches (i.e., Decision Trees, Bayesian Network, Naive Bayes, Logistic Regression) to predict the performance of a testing technique. During online learning, they adapt the selection of test cases to the data observed as the testing proceeds. Yu et al. \cite{yu2015does} proposed a technique that can be used to distinguish failing tests that executed a single fault from those that executed multiple faults. The technique combines information from a set of fault localization ranked lists, each produced for a certain failing test, and the distance between a failing test and the passing test that most resembles it.

\subsubsection{Testing Applications}

Malik et al. \cite{malik2013automatic} applied three unsupervised approaches and one supervised approach to detect performance deviations. They first generated performance signatures which are
minimal sets of performance counters, describing the essential characteristics of a System Under Test (SUT) for a given load test. They then classified each observed signature into passed or failed, and conducted a root-cause analysis of detected deviations. To reduce expensive calculation of mutation testing, Zhang et al. \cite{zhang2018predictive} proposed predictive mutation testing (PMT), that can predict the testing results without executing mutants. After identifying execution features, infection features and propagation features of tests and mutants, they selected Random Forest algorithm to construct a classification model for predicting whether a mutant can survived or be killed without mutant execution.

%that are minimal sets of performance counters that describe the essential characteristics of a System Under Test (SUT) for a given load test

\subsubsection{Software Defect Prediction}

Software defect prediction techniques are employed to help prioritize software testing and debugging. These techniques can recommend software components that are likely to be defective to developers. Rahman et al. compared static bug finders and defect predictive models \cite{rahman2014comparing}. They found that in some cases, the performance of certain static bug-finders can be enhanced using information provided by statistical defect prediction models. Therefore, applying prediction models to defect prediction has become a popular research direction.

Lessmann et al. \cite{lessmann2008benchmarking} proposed a framework for comparative software defect prediction experiments. They conducted a large-scale empirical comparison of 22 classifiers over 10 public domain data sets. Their results indicated that the importance of the particular classification algorithm may be less than previously assumed. This is because no significant performance differences could be detected among the top 17 classifiers. However, Ghotra et al. \cite{ghotra2015revisiting} doubted this conclusion and they pointed that the datasets Lessmann et al. used were both noisy and biased. Therefore, they replicated the prior study with initial its datasets as well as new datasets after cleansing. Their new experimental results demonstrated that some classification techniques were more suitable for building predictive models. They showed that Logistic Model Tree when combined with ensemble methods (i.e., bagging, random subspace, and rotation forest) achieves top-rank performance. Furthermore, clustering techniques (i.e., Expectation Maximization and K-means), rule-based techniques (Repeated Incremental Pruning to Produce Error Reduction and Ripple Down Rules), and support vector machine perform worse than other predictive models. Herzig et al. \cite{herzig2016impact} conducted an analysis of the impact of tangled code changes in defect prediction. They used a multi-predictive model to identify tangled changes and they found that untangling tangled code changes can achieve significant accuracy improvements on defect prediction.

Many predictive model algorithms have several tunable parameters and their values may have a large impact on the model's prediction performance. Song et al. \cite{song2010general} proposed and evaluated a general framework for defect prediction that supports unbiased and comprehensive comparison between competing prediction systems. They first evaluated and chose a good learning scheme, consisting of a data preprocessor, an attribute selector and a learning algorithm. They then used the scheme to build a predictor. Their framework proposes three key elements for defect prediction models, i.e., datasets, features and algorithms.  Tantithamthavorn et al. \cite{tantithamthavorn2016automated} conducted a case study on 18 datasets to investigate the performance of an automated parameter optimization technique, Caret, for defect prediction. They tried 26 classification techniques that require at least one parameter setting and concluded that automated parameter optimization techniques like Caret yield substantially benefits in terms of performance improvement and stability, while incurring a manageable additional computational cost.

Features, metrics or attributes are crucial for the building of a well-performed defect predictive model. A number of studies researched a lot about feature extraction for defect prediction \cite{rahman2013and, shivaji2012reducing, wang2016automatically}. Kim et al. \cite{kim2008classifying} introduced a new technique for change-level defect prediction by using support vector machine. They are the first to classify file changes as buggy or clean leveraging change information features.  An obvious advantage of change-level classification is that predictions can be performed immediately upon the completion of changes. Shivaji et al. \cite{shivaji2012reducing} investigated multiple feature selection techniques that are generally applicable to classification-based bug prediction methods. They found that binary features are better, and that between 3.12\% and 25\% of the total feature set yielded optimal classification results. Lee et al. \cite{lee2011micro} proposed 56 new micro interaction metrics (MIMs) by using developers' interaction information (e.g., file editing and selection events) in an Eclipse plug-in, Mylyn. To evaluate the effectiveness of MIMs, they built defect predictive models using traditional metrics, MIMS and their combinations. The evaluation results showed that MIMs allowed the performance of those models improve significantly.  Yu et al. \cite{yu2018conpredictor} initially focused on defect prediction for concurrent programs and proposed ConPredictor, a prototype tool that used four classification techniques to identify defects by applying a set of static and dynamic code metrics based on unique features of concurrent programs. Lin et al. \cite{lin2019identifying} present an approach to identifying buggy videos from a set of gameplay videos. They applied three classification models -- logistic regression, neural network and random forests -- to determine the probability that a video showcases a bug. They observed that random forests achieved the best performance among classifiers. Afterwards, they identified key features in game videos and used random forest classifier to prioritize game videos according to the likelihood of each video containing bugs.

A good predictive model relies heavily on the dataset it learns from, which is also the case for models for software defect prediction. Many studies investigate the quality, such as bias and size, of datasets for defect prediction. Bird et al. \cite{bird2009fair} used predictive models to investigate historical data from several software projects (i.e., Eclipse and AspectJ), and found strong evidence of systematic bias. Rahman et al. \cite{rahman2013sample} investigated the effect of size and bias of datasets on performance of defect prediction using logistic regression model. They investigated 12 open source projects and their results suggested that the type of bias has limited impact on prediction results, and the effect of bias is strongly confounded by size. Tantithamthavorn et al. \cite{tantithamthavorn2015impact} used random forest to investigate the impact of mislabelling on the performance and interpretation of defect models. They found that precision is rarely impacted by mislabelling while recall is impacted much by mislabelling. Moreover, the most influential features are generally robust to mislabelling.

For datasets of poor quality, researchers also have proposed several approaches to address their issues. Kim et al. \cite{kim2011dealing} proposed an approach, named CLNI, to deal with the noise in defect prediction datasets. CLNI can effectively identify and eliminate noise. The noise eliminated from the training sets produced by CLNI was shown to improve the defect prediction model's performance. Nam et al. \cite{nam2015clami} proposed novel approaches CLA and CLAMI, which can work well for defect prediction on unlabeled datasets in an automated manner without any manual effort. Gong et al. \cite{gong2019empirical} investigated the impact of class overlap and class imbalance problems on defect prediction, and then present an improved approach (IKMCCA) to solve those problems in order to improve defect prediction performance for within project defect prediction (WPDP) and cross project defect prediction (CPDP). Tantithamthavorn et al. \cite{tantithamthavorn2018impact, mori2019balancing} conducted an empirical study to investigate how class rebalancing techniques impact the performance and interpretation of defect prediction. They also explored experimental settings that can help rebalancing techniques achieve the best performance in defect prediction. They concluded that using different metrics and classification algorithms allowed the performance of predictive models to vary, indicating that researchers should avoid class rebalancing techniques when deriving understandings and knowledge from models.

%• The impact of class rebalancing techniques on the interpretation of defect prediction models relies heavily on the used classification techniques, suggesting that researchers and practitioners should avoid class rebalancing techniques when deriving knowledge and understandings from defect prediction models.

Datasets from many studies are unfortunately not always available due to  privacy policies and other factors. To address the privacy problem, Peters et al. \cite{peters2012privacy, peters2013balancing, peters2015lace2}  measured the utility of privatized datasets empirically using Random Forests, Naive Bayes and Logistic Regression. Through this they showed the usefulness of their proposed privacy algorithm MORPH \cite{peters2012privacy}. MORPH is a data mutator that moves the data a random distance, while not across the class boundaries. In a later work \cite{peters2013balancing}, they improved MORPH by proposing CLIFF+MORPH to enable effective defect prediction from shared data while preserving privacy. CLIFF is an instance pruner that deletes irrelevant examples. Recently, they again extended MORPH to propose LACE2 \cite{peters2015lace2}.

\textbf{Deep learning based defect prediction:}
Recently, deep learning, as an advanced machine learning algorithm, has become widely discussed and applied including in Software Engineering. Several researchers have tried to improve the performance of defect prediction via deep learning \cite{jing2014dictionary, wang2016automatically}. Jing et al. \cite{jing2014dictionary} proposed a cost-sensitive discriminative dictionary learning (CDDL) approach for software defect prediction. CDDL is based on sparse coding which can transform the initial features into more representative code. Their results showed that CDDL is superior to five representative methods, i.e., support vector machine, Compressed C4.5 decision tree, weighted Naive Bayes, coding based ensemble learning (CEL), and cost-sensitive boosting neural network. Wang et al. \cite{wang2016automatically} leveraged Deep Belief Network (DBN) to automatically learn semantic features from token vectors extracted from programs Abstract Syntax Trees. Their evaluation on ten open source projects showed that learned semantic features significantly improve both within-project defect prediction and cross-project defect prediction compared to traditional features.

%\textbf{Defect prediction model features:}

%\textbf{Defect prediction model datasets:}

\textbf{Cross project defect prediction:} Cross-project defect prediction is a topic of growing research interest. It uses data from one project to build the predictive model and predicts defects in another project based on the trained model so that it can solve the problem that there is no sufficient amount of data available to train within a project, such as a new project. Some studies concentrated on cross-project defect predictive models on a large scale \cite{zimmermann2009cross}.

To improve the performance of cross-project defect prediction, researches have tried several techniques \cite{jing2015heterogeneous, nam2017heterogeneous, sun2016inteq}. Nam et al. \cite{nam2013transfer} proposed a novel transfer defect learning approach, TCA+, extending a transfer learning approach Transfer Component Analysis (TCA) . TCA+ can provide decision rules to select suitable normalization options for TCA of a given source-target project pair. In a later work, they addressed the limitation that cross-project defect prediction cannot be conducted across projects with heterogeneous metric sets by proposing a heterogeneous defect prediction approach. Jiang et al. \cite{jing2015heterogeneous} proposed an approach CCA+ for heterogeneous cross-company defect prediction. CCA+ combines unified metric representation and canonical correlation analysis and can achieve the best prediction results with the nearest neighbor classifier.

Zhang et al. \cite{zhang2016cross} found that connectivity-based unsupervised classifiers (via spectral clustering) offer a viable solution for cross-project defect prediction. Their spectral classifier ranks as one of the top classifiers among five widely-used supervised classifiers and five unsupervised classifiers in cross-project defect prediction. Due to the impact of heterogeneous metric sets, current defect prediction techniques are difficult to use in CPDP. To address this problem, Nam and Kim \cite{nam2017heterogeneous} trained a predictive model for heterogeneous defect prediction (HDP) with heterogeneous metric sets by applying metric selection and metric matching techniques. Zhou et al. \cite{zhou2018far} noticed that most CPDP models were not compared against simple defect predictive models, and thus they aimed to investigate whether CPDP models really performed well compared against simple models, ManualDown and ManualUp. To their surprise, they found that the performance of ManualDown and ManualUp were superior to most of existing CPDP models.

\textbf{Just-in-time (JIT) defect prediction:} Compared with traditional defect predictions at class or file level, Just-in-Time (JIT) defect prediction is of more practical value for participants, which aims to identify defect-inducing changes. Many studies focused on JIT defect prediction by employing SZZ approach \cite{fan2019impact, shivaji2012reducing, kamei2016studying, kamei2012large}. In order to identify bug-introducing changes, SZZ first detects the bug-fixing changes whose the change log contains bug identifier. Among the bug-fixing changes, SZZ identifies the modified the buggy lines so that the bug can be removed. SZZ then traces the code change history to search for potential bug-introducing changes that may introduce buggy lines. Finally, the remaining changes are deemed as bug-introducing after SZZ eliminates the incorrect ones from the initial set of potential bug-introducing changes. Yang et al. \cite{yang2016effort} used the most common change metrics from various source to build unsupervised and supervised models to predict defect-inducing changes. Comparing these models under cross-validation and time-wise-cross-validation, their results showed that the simple unsupervised models performed better than state-of-the-art supervised models for JIT defect prediction.

\textbf{Failure prediction:}
%Failure prediction can be split into offline application and online application by analyzing the time horizon.
Failure prediction is critical to predictive maintenance since it has the ability to prevent maintenance costs and failure / bug occurrences. Salfner et al. \cite{salfner2010survey} conducted a survey on online failure prediction approaches. Online failure prediction is based on runtime monitoring and a variety of models and methods that use the current state of a system and, frequently, the past experience as well. This study classified the existing online failure prediction techniques into three main categories: failure tracking-based, error reporting-based, and symptom monitoring-based approaches. Yilmaz and Porter \cite{yilmaz2010combining} classify measured executions into successful and failed executions in order to apply the resulting models to systems with an unknown failure status. They further applied their technique to online environments. Ozcelik and Yilmaz \cite{ozcelik2015seer} proposed a online failure prediction approach for combining hardware and software instrumentation. In the training phase, they applied global and frequency filtering to pick up candidate functions from historical data as the input of the proposed approach and created a predictive model for identifying the best performing functions that best distinguish failing executions from passing executions. These best performing functions were referred to as seer functions. Then their proposed model makes a binary prediction for each seer function into passing or failing during monitoring phase. To address the understudied problem of inferring concurrency-related documentations, Habib et al. \cite{habib2018class} developed a novel tool, TSFinder, to automatically classify classes as thread-safe or thread-unsafe with a combination of static analysis and graph-based classification. Yilmaz and Porter \cite{yilmaz2010combining} present a hybrid instrumentation approach to distinguish failed executions from successful executions with consideration of potential cost-benefit tradeoffs. After collecting program spectra and augmenting the underlying data with low cost, they trained a classification model with J48 algorithm to classify program executions.

\subsubsection{Bug Report Management}
%Bug reports are vital for any software development. They allow users to inform developers of the problems encountered while using a software. Bug reports typically contain a detailed description of a failure and occasionally hint at the location of the fault in the code (in form of patches or stack traces).

Bug reports are essential for any software development, which typically contain a detailed description of the failure and occasionally hint at the location of the fault in the source code in the form of patches or stack traces. They also allow users to inform developers of problems encountered while applying software applications.

\textbf{Duplicate bug report detection:}
To reduce redundant effort, there has been much research into identification of duplicate bug reports \cite{sun2010discriminative, sun2011towards}. Sun et al. \cite{sun2010discriminative} used discriminative models to identify duplicates more accurately. They used a support vector machine (SVM) with linear kernel based on 54 text features. In a later work \cite{sun2011towards}, they more fully utilized the information available in bug reports including not only textual content and description fields, but also non-textual fields (e.g., product, component, version) and proposed a retrieval function REP to identify duplicated reports. A two-round stochastic gradient descent was applied to automatically optimize REP in a supervised learning manner.

\textbf{Bug report quality management:}
Zanetti et al. \cite{zanetti2013categorizing} proposed an efficient and practical method to identify valid bug reports i.e., the bug reports that refer to an actual software bug, are not duplicates, and contain enough information to be processed right away. They used support vector machine to identify valid bug reports based on nine network measures using a comprehensive data set of more than 700,000 bug reports obtained from the BUGZILLA installation of four major OSS communities. To effectively analyze anomaly bug reports, Lucia et al. \cite{lo2012active} proposed an approach to automatically identify and refine bug reports by the incorporation of user feedback. They first listed the top few anomaly reports from the list of reports generated by a tool in its default ordering. Based on feedback of users who either accepted or rejected each of the reports, their tool adopted nearest neighbor with non-nested generalization (NNGe) to automatically and iteratively refine a classification model for anomalies and resorted the rest of the reports.

\textbf{Bug report assignment and categorization:}
Considering the impact of bug report assignment on the efficiency of development process, making a proper and effective assignment of each bug report is a crucial task. This can save much time and effort on bug-fixing activity. Jeong et al. \cite{jeong2009improving} introduced a graph model based on Markov chains, which captures bug tossing history. They showed that the accuracy of bug assignment prediction is improved using naive bayes with tossing graph. Anvik et al. \cite{anvik2011reducing} presented a machine learning approach to create recommenders that assist with a variety of decisions aimed at streamlining the development process. They built three different development-oriented recommenders by applying six predictive models  to suggest which developers might fix errors in the bug reports, which product components a report might pertain to, and which developers on the project might be interested in following the report. Jonsson et al.\cite{jonsson2016automated} conducted a study to evaluate machine learning classification based automated bug assignment techniques. They combined an ensemble learner Stacked Generalization (SG) with several classifiers and evaluated their performance with over 50,000 bug reports from two companies. Peters et al. \cite{peters2017text} present a framework to filter and remove non-security bug reports containing security related keywords by training five predictive models -- Random Forest, Naive Bayes, Logistic Regression, Multilayer Perceptron, and K Nearest Neighbor -- on Four Apache projects and Chromium involving 45,940 bug reports. They found that their framework mitigates the class imbalance issue and reduces the number of mislabelled security bug reports by 38\%. Proper bug report assignment also relies on proper bug report categorization. There are a number of studies that propose techniques to categorize bug reports. Among them a popular research area is reopened bug report prediction. Zimmermann et al. \cite{zimmermann2012characterizing} characterized and predicted which bugs get reopened. They first qualitatively identified causes for bug reopens, and then built logistic regression model to predict the probability that a bug will be reopened.

%Xia et al. \cite{xia2015automatic} proposed a novel approach called ReopenPredictor, which extract more textual features from the bug reports and combines decision tree and multinomial Naive Bayes to yield better performance for reopened bug prediction.

\textbf{Bug report-based bug fixing tasks:}
A number of studies leverage bug report contents to address other bug fixing related tasks. Guo et al. \cite{guo2010characterizing} performed an empirical study to characterize factors that affect which bugs get fixed in Windows Vista and Windows 7. They concentrated on investigating factors related to bug report edits and relationships between people involved in fixing bugs, and built a statistical model using logistic regression to predict the probability that a new bug will be fixed. Kim et al. \cite{kim2013should} proposed a two phase predictive model to identify the files likely to be fixed by analyzing bug reports contents. In order to collect predictable bug reports, they checked whether the given bug report contains sufficient information for prediction (predictable or deficient) in the first phase.  Zhang et al. \cite{zhang2013predicting} proposed a Markov-based method for predicting the number of bugs that will be fixed in future. For a given bug report, they also constructed a KNN classification model to predict bug-fixing effort based on the intuition of similar bugs requiring similar time cost.

%Murgia et al. \cite{murgia2018exploratory} conducted an analysis on whether issue reports, as a common software artifact, conveyed developers' emotional information. They trained a classifier to identify issue reports containing different emotions by using machine learning techniques. Through an analysis of experimental results, they observed that humans were easier to agree on some emotions, such as gratitude, joy and sadness. However, they started to doubt their interpretation when more contexts of an issue report were provided.

\subsubsection{Software Quality Assessment}

With the continuous release of new software versions, developers have been committed to maintaining a high level of software quality during the whole of SDLC. Generally, Software quality can be defined as ``the degree to which a software product meets established requirements" and it involves many aspects, such as desirable characteristics of software products and process, techniques, or tools used to achieve those characteristics. In this section, we introduce the studies using predictive models for assessing software quality.

\textbf{Vulnerability Detection:}
An indicator being commonly used for evaluating software quality is software vulnerability. There are many studies on vulnerability detection using predictive models \cite{shin2010evaluating, shar2013mining, scandariato2014predicting, zhang2019machine, dam2018automatic}. Shin et al. \cite{shin2010evaluating} conducted an empirical case study to investigate the impact of software metrics obtained in the early SDLC on predicting vulnerable code. They collected complexity, code churn, and developer activity metrics from two large-scale projects (Mozilla Firefox and the Linux kernel) and trained five machine learning models to identify vulnerabilities by applying these metrics. The experimental results showed that Naive Bayes achieve the best performance among other classifiers. Scandariato et al. \cite{scandariato2014predicting} present an approach to predict which components of a software application contain security vulnerabilities by training five machine learning models. Based on \cite{shar2013mining}, Zhang et al. \cite{zhang2019machine} utilized deep learning techniques to detecting SQL injection vulnerabilities in PHP code. They observed that a classifier trained using CNN outperformed another one trained by Multilayer Perceptron (MLP).  Dam et al \cite{dam2018automatic} built a Long Short Term Memory deep learning-based model to automatically identify key features that predict vulnerable software components. Their experiments on Firefox and 18 Android apps showed that LSTM performs better than other predictive models for both within-project and cross-project vulnerability detection.

%\subsubsection{Malware Detection}
%Saccente et al. \cite{saccente2019project} developed a prototype tool for identify method-level vulnerabilities within source code. The tool employed various data preparation methods to be independent of coding style and to automate the process of extracting methods, labeling data, and partitioning datasets.

%They are likely to implement functionalities which contradict user and organisational interests \cite{aslan2020comprehensive}.
\textbf{Malware detection:}
Some software, especially mobile apps, can include significant malware.  Generally, malware includes viruses, worms, trojans, key loggers and spyware, and are harmful at diverse severity scales. Malware can lead to damages of varying severity, ranging from spurious app crashes to financial losses with malware sending premium-rate SMS, and to private data leaks. A number of studies have aimed to better detect malware by using predictive models \cite{chandramohan2013scalable, avdiienko2015mining}.

Chandramohan et al. \cite{chandramohan2013scalable} proposed and evaluated a bounded feature space behavior modeling (BOFM) framework for scalable malware detection. They first extracted a feature vector that is bounded by an upper limit N using BOFM, and then trained an SVM to detect malware. Avdiienko et al. \cite{avdiienko2015mining} compared data flow of 2,866 benign applications against those in malicious applications to capture differences. They noticed that malicious applications treat sensitive data differently than benign applications, which can be used to identify malicious applications. Therefore, they present MUDFLOW prototype to flag malware by using abnormal data flow.  Since there is no studies to detect malicious and aggressive push notification advertisement, Liu et al. \cite{liu2019dapanda} provided a taxonomy of push notifications and present a crowdsourcing-based approach to automatically detect aggressive push notifications in Android apps. Their approach used a guided testing approach to record push notifications and flagged the malicious ones by analyzing runtime information.

\textbf{Product Quality Assessment:}
High quality software products can improve user experience and reduce developers' effort in  maintenance. Quality assessment is an important task to discover and fix new bugs that appear in products timely. Müller et al. \cite{muller2016using} investigated whether biometrics are able to determine code quality concerns. They conducted a field study with ten professional developers and further adopted machine learning classifiers to predict developers' perceived difficulty of code elements while working on a change task. 
%Experimental results showed that biometrics are indeed able to predict quality concerns of parts of the code, and lower development and evaluation costs, improving upon a naive classifier by more than 26\% and outperforming classifiers based on traditional metrics. 
Mills et al. \cite{mills2017predicting} proposed a solution to the problem of predicting query quality by training a machine learning classifier based on 28 query measures. They applied a Random Forest algorithm to 12 open source systems implemented in Java and C++, defining seven post-retrieval query quality properties, which extended 21 pre-retrieval properties proposed by previous work \cite{haiduc2012evaluating, haiduc2013automatic}. In order to improve the efficiency of developers in handling test reports, Chen et al. \cite{chen2020systemic} present a framework to assess the quality of crowdsourced test reports. They trained a classifier using logistic regression to predict the quality of reports. The proposed framework achieved good performance in terms of precision, recall and F-measure.

\subsection{Software Maintenance}

Software maintenance is an integral stage of the software life cycle. No software product is ever completely finished and all need some form of ongoing maintenance. We introduce representative applications of predictive models for software maintenance.

\subsubsection{Configuration Management}

When developers or researchers build predictive models in different application scenarios, the effect of model configuration, i.e., the parameters, specify their behavior. There are several studies focusing on the effect of parameters in predictive models or which parameter values lead to the best performance \cite{thomas2013impact, song2013itree, DBLP:conf/icse/SongPF12}.

Thomas et al. \cite{thomas2013impact} investigated the effectiveness of a large space of classifier configurations, 3,172 in total, and present a framework for combining the results of multiple classifier configurations since classifier combination has shown promise in certain domains. 
%The evaluation results demonstrated that parameters of a classifier has a significant impact on the model's performance. They found that combining multiple classifiers improves the performance of even the best individual classifiers.
 Due to several limitations of the combinatorial interaction testing (CIT) approach, Song et al. \cite{song2013itree} implemented an iterative learning algorithm called iTree, which effectively searched for a small part of configurations that closely approximated the effective configuration space of a system. Based on their previous work \cite{DBLP:conf/icse/SongPF12}, the key improvements of  are based on the use of composite proto-interactions, i.e. a construct that improves iTrees ability to correctly learn key configuration option combinations. This in turn significantly improves iTrees running time, without sacrificing effectiveness. Nair et al. \cite{nair2017using} present a rank-based approach to rank software configurations and identify the optimal value without requiring exact performance values. Experiments were conducted on 21 scenarios for evaluating effectiveness of their strategy, and results showed that rank-based approach allows building an accurate performance model with very few data instances, helping developers to significantly reduce the cost of building models.

\subsubsection{Software Change Management}

Change management involves tracking and managing changes to artifacts, such as requirements, change requests, bug reports, source code files, and other digital assets. It's critical for effective application development. With the rapid evolution and change of software, effective change management has become a major challenge in the industry, and thus many studies have worked on identifying changes, analyzing changes, tracing changes, assessing security or metrics using predictive models \cite{falessi2018leveraging}.
%\textbf{Software Change analysis:}

 Padhye et al. \cite{padhye2014needfeed} present a prototype tool for highlighting code changes in order to improve development efficiency. They explored the use of various techniques including Naive approach (TOUCH-based subscription strategy) and machine learning algorithms (Conjunctive Rule, J48 and Naive Bayes) to model code relevance. Their proposed tool can highlight changes that developers may need to review in order to personalize a developers' change notification feed. 
 %They found that the best strategy can reduce the notification clutter by more than 90\%. They then applied these methods at a different granularity.

Liu et al. \cite{liu2019predicting} proposed a novel learning-based approach that can predict software licenses as software changes by applying linear-chain conditional random field (CRF) learning algorithm. To demonstrate its effectiveness, they evaluated their approach on 700 open source projects.
% The experimental results show the performance of their approach significantly surpasses the performance of three baselines, involving a state-of-the-art license prediction tool, Ninka. 
Misirli et al. \cite{misirli2016studying} analyzed high impact fix-inducing changes (HIFCs). They used a Random Forest-based algorithm to identify HIFCs and determine the best indicators of HIFCs on six large open source projects. They also present a measure to evaluate the impact of fix-inducing changes. Falessi et al. \cite{falessi2018leveraging} introduced new Requirements to the Requirements Set (R2RS) family of metrics based on some intuitions. In order to evaluate the usefulness of the proposed metrics, they applied five classifiers to predict the set of classes impacted by a requirement over 700,000 classes. 
Horton and Parnin et al. \cite{horton2019v2} proposed V2, a strategy to detect discrete instances of configuration drift in order to determine whether a code snippet is out-of-date and where code snippets that use APIs experience a breaking change. %To reduce the number of potential environment configurations without validation, V2 explored the possible configuration research space for a code segment by adopting feedback-directed search.

\subsubsection{Code Clone Management}

Code clones have always been a double-edged sword in software development. On one hand, developers reuse the existing code snippets to complete development tasks, largely reducing coding effort. On the other hand, creating clones in source code may introduce new defects or bugs, which can lead to extra maintenance effort to ensure consistency among cloned code snippets. Since software clones have a clear influence on software maintenance and evolution, many studies \cite{wang2012can, wang2014predicting,  saini2018oreo, nafi2019clcdsa, mostaeen2019clonecognition, thongtanunam2019will} proposed various approaches to detect and management code clones in source code.

\textbf{Code Clone Detection:} Saini et al. \cite{saini2018oreo} implemented a novel code clone detection approach referred as to Oreo, which can be capable of detecting harder-to-detect clones (i.e., semantic clones) in the Twilight Zone. They employed five different models to detect clones and selected siamese architecture neural network as the final choice. They evaluated the recall of Oreo onvBigCloneBench, and Oreo achieved both high recall and precision. Nafi et al. \cite{nafi2019clcdsa} proposed a framework which can detect cross language clones (CLCs) with no need to generate the intermediate representation of source code. Their framework analyzed the different syntactic features of source code across different programming languages and applied siamese architecture neural network to detect CLCs. %Their framework outperformed  state-of-the-art approaches in detecting cross language clones in terms of precision, recall, and F-measure.

\textbf{Code Clone Analysis:} Prior studies had demonstrated that neither the percentage of harmless code clones nor that of harmful code clones is negligible \cite{gode2011frequency}. In order to assist developers in avoiding the negative impacts of harmful code clones and taking advantage of the benefits of clones, they present a novel approach that can automatically predict the harmfulness of an intended cloning operation after performing copy-and-paste by using Bayesian Networks. 
%They used two large-scale industrial software projects to evaluate their approach under two usage scenarios: conservative scenario and aggressive scenario. The empirical results showed that their approach had the good performance in both scenarios. 
Since only a small part of code clones experience consistent changes during software evolution history, Wang et al. \cite{wang2014predicting} defined a code cloning operation as consistency-maintenance-required if a code clone experience consistent changes in their evolution history, and leveraged Bayesian Networks to automatically predict whether a code cloning operation needs consistency maintenance when developers performing copy-and-paste operations. They evaluated the effectiveness of their approach on four projects under two usage scenarios. In order to perform the efficiency of clone management efforts, Thongtanunam et al. \cite{thongtanunam2019will} conducted an empirical study on six open-source Java systems to investigate the life expectancy of clones. They found that 30\% to 87\% of clones were short-lived, and these short-lived clones were changed more frequently than long-lived clones throughout their lifetime. They also applied Random Forest classifier to predict the life expectancy of newly-introduced clones. 
%Furthermore, they noticed that several features can influence the life expectancy of newly-introduced clones, such as the size and complexity of clones. Since many clone detectors return code fragments that are not considered clones by users, it is necessary to perform manual validation of the reported possible clones.  
Mostaeen et al. \cite{mostaeen2019clonecognition} present a tool called CloneCognition that can automate the laborious manual validation process by using Artificial Neural Network (ANN). Their tool showed promising performance when compared with state-of-the-art clone validation techniques.

\subsubsection{Code Smell Detection}

Code smell refers to the symptoms of poor design and implementation choices in source code, which possibly indicate deeper problems for further development, maintenance, and evolution of software. A large number of studies have been proposed to automatically detect various types of code smells since it is tedious and time consuming to manually identify code smells \cite{mens2004survey}.

Fontana et al. \cite{fontana2016comparing} applied 16 different machine learning techniques on large-scale code smell instances. By comparing the performance of various machine learning algorithms, they found that all algorithms performed well on cross-validation datasets. Their experimental results showed that the highest performance was achieved by J48 and Random Forest, while SVMs had the worst performance in code smell detection. Palomba et al. \cite{palomba2018beyond} performed an empirical study to investigate the relationship between community smells and code smells. 
%Community smells reflect the presence of organizational and socio-technical issues within a software community that may lead to additional project costs. Prior studies provided pieces of evidence that are often connected to circumstances such as code smells. Thus they proposed a code smell intensity predictive model and conducted a fine-grained analysis on 117 datasets for exploring how and the extent to which community smells impact code smell intensity. 
They found that community smells result in the increasing intensity of code smells in projects. Liu et al. \cite{liu2018deep} proposed a neural network-based classifier that detects feature envy without the need for manually selecting features. Evaluation results showed that the proposed approach significantly improved the state-of-the-art in identifying feature envy smells.

%community smells reflect the presence of organizational and socio technical issues within a software community that may lead to additional project costs. Recent empirical studies provide evidence that community smells are often—if not always—connected to circumstances such as code smells.

\subsubsection{API Issue Classification}

In order to identify API issue-related sentences in Stack Overflow (SO) Posts and to classify SO posts concerning API issues, Lin et al. \cite{lin2019pattern} introduced a machine learning based approach called POME. POME classifies sentences related to APIs in Stack Overflow, using natural language parsing and pattern-matching and determined their polarity (positive vs negative).  Evaluation results showed that POME exhibited a higher precision compared with a state-of-the-art approaches. Similarly, Ahasanuzzaman et al. \cite{ahasanuzzaman2020caps} developed a supervised learning approach with a Conditional Random Field (CRF). They also performed an investigation to identify important features and test the performance of the proposed technique for classifying issue sentences. Petrosyan et al. \cite{petrosyan2015discovering} classified fragmented tutorial sections that help developers to understand a subset of API types. They applied a supervised text classification based on structural and linguistic features. applying supervised text classification according to structural and linguistic features. Nam et al. \cite{nam2019marble} conducted studies on boilerplate code and investigated what properties make code be considered as ``boilerplate". They then proposed a novel approach automatically mine boilerplate code candidates from API client code repositories.

\subsubsection{Log Analysis}

Logging provides visibility into the health and performance of an application and infrastructure stack. Some studies concentrated on log analysis in order to enable development teams and system administrators to more easily diagnose and rectify issues. Russo et al. \cite{russo2015mining} proposed an approach for classification and prediction of defective log sequences. They used three well-known SVMs -- radial basis function, multilayer perceptron and linear kernels -- to predict and fit log sequences containing defects with high probability. The approach achieves comparable accuracy as other log analysis techniques. Li et al. \cite{li2017towards} studied reasons of log changes and proposed an approach to determine whether log change suggestions are required while committing a code change. Zhang et al. \cite{zhang2019robust} present LogRobust, a log-based anomaly detection approach, using an attention-based Bi-LSTM model. They evaluated LogRobust on Hadoop and the results demonstrated that LogRobust can identify and handle stable log events and sequences with high accuracy.

\subsubsection{Code Comment Analysis}

Comments in source code explain the purpose of the code and help with code understanding during development and maintenance. Huang et al. \cite{huang2018identifying} present a classification approach to detect self-admitted technical debt (SATD) in code comments. They applied feature selection to select valuable features and trained a composite classifier by combining multiple classifiers for identifying SATD comments in open source projects. %Evaluation results showed that the proposed approach outperformed the state-of-the-art and baselines in terms of precision, recall and F-measure. 
For better understanding the goal and target audience of code comments, Pascarella et al. \cite{pascarella2019classifying} investigated how diverse Java software projects used code comments, and applied machine learning algorithms to automatically classify code comments at line level.

\subsubsection{Software Repository Mining}

Software repositories, such as GitHub and Q\&A sites, contain a large amount of knowledge related to software development. Mining these repositories is able to recover valuable knowledge that can help developers address difficulties when programming. However, the quality of a high proportion of questions and answers in repositories is still a concern. To ensure that researchers reach realistic and accurate conclusions, Munaiah et al. \cite{munaiah2017curating} trained a Random Forest classifier for sieving out the noise in GitHub repositories. They evaluated their framework on 200 repositories with known ground truth classification. Xu et al. \cite{xu2016predicting} applied deep learning techniques to predict Semantically Linkable Knowledge units in Stack Overflow. They formulated the problem of identifying knowledge units as a multi-class classification task and trained a convolutional neural network (CNN) model by exploiting informative word-level and document-level features. Prana et al. \cite{prana2019categorizing} manually annotated 4,226 README files from GitHub repositories and designed a multi-label classifier to classify these files into eight different categories.

\subsubsection{App Review Mining}

In order to capture the meaning of a sentence at a higher level of abstraction, Jha and Mahmoud \cite{jha2018using} applied frame semantics to generate low-dimensional representations of text and used SVM and Naive Bayes to classify app store reviews into three categories of actionable software maintenance requests (i.e., bug reports, feature requests, and otherwise), enhancing the predictive capabilities and reducing the chances of overfitting. Martens et al. \cite{martens2019towards} performed a survey of 43 fake review providers to study the significant differences between fake reviews and non-fake reviews. They then implemented seven classifiers to automatically detect fake reviews in app stores. The experimental results showed that the Random Forest algorithm achieved the best performance compared to other models.

%various categories of actionable software maintenance requests. to capture the meaning of a sentence at a higher level of abstraction

\subsection{Software Engineering Professional Practice}

Software engineering professional practice is related to the skills, knowledge and attitudes that software engineers need to possess to practice software engineering in a professional, responsible and ethical manner. The study of professional practices includes the areas of technical communication, group dynamics and human aspects including psychology, emotions, team dynamics, social and professional responsibilities \cite{shafer2014csdp}. On the other hand, the goal of software development is to meet user desires or needs \cite{bourque2014guide}. We introduce some key related studies into these developers' and users' perspectives that use predictive models.

%Thus it not only takes into consideration professional practice, but also various human users' factors when developing products \cite{bourque2014guide}.

\subsubsection{Developer Behavior Analysis}

Meneely et al. \cite{meneely2011does} performed an empirical and longitudinal case study of a large Cisco networking product over a five year period. They examined statistical correlations between monthly team-level metrics and monthly product-level metrics. Their linear regression predictive model based on team metrics was able to predict the products post-release failure rate within a 95\% prediction interval for 38 out of 40 months. Canfora et al. \cite{canfora2012going} proposed an approach named Yoda, that aimed to identify and recommend mentors in software projects by mining data from mailing lists and versioning systems. Fritz et al. \cite{fritz2014using} implemented a novel approach to detect when software developers are experiencing difficulty while they work on their programming tasks, and stop developers' behavior before they introduce bugs into the code. They classify the difficulty of code comprehension tasks using data from psycho-physiological sensors and chose Naive Bayes as the classification algorithm because its training can be updated on-the-fly.

Muller et al. \cite{muller2015stuck} investigated developers emotions, progress and applied biometric measures to classify them in the context of software change tasks. They trained a J48 decision tree to distinguish between positive and negative emotions based on biometric measurements. Bacchelli et al. \cite{bacchelli2012content} presented an approach to classify email content at line level. Their technique fused naive Bayes algorithm with island parsing to perform automatic classification of the content of development emails into five language categories: natural language text, source code fragments, stack traces, code patches, and junk. Their technique can help developers subsequently apply ad hoc analysis techniques for each category. Later, Sorbo et al. \cite{di2015development} proposed a semi-supervised approach named DECA (Development Emails Content Analyzer) to mine intention from developer emails. DECA uses Natural Language Parsing to classify the content of development emails according to their different purposes (e.g., feature request, opinion asking, problem discovery, solution proposal, or information giving), identifying email elements that can be used for specific tasks. They showed the superiority of DECA to traditional machine learning techniques.

Zhou et al. \cite{zhou2014will} measured, understood, and predicted how the newcomers involvement and environment in  issue tracking systems affects their odds of becoming a long term contributor. They constructed nine measures of involvement and environment based on events recorded in an issue tracking system and used logistic regression model to predict long term contributors. Wood et al. \cite{wood2018detecting} conducted an empirical study with 30 professional programmers and trained a supervised learning algorithm to identify speech act types in developers' conversations in order to obtain useful information for bug repair.

\subsubsection{User Behavior Analysis}

Murukannaiah et al. \cite{murukannaiah2015platys} believed that an application cannot benefit users by employing location information, but rather a position, the abstraction of location, such as restaurant, park, home, and office. They referred this level of location abstraction to as ``place". Therefore, they proposed the Platys framework as a way to address the special challenges of place-aware application development. They collected place labels and Android phone sensor readings from 10 users, and applied Platys to learn each users places by combining active learning and semi-supervised learning. Zheng et al. \cite{zheng2019ifeedback} implemented a tool referred to as iFeedback to perform real-time issue detection by analyzing user feedback texts. iFeedback trained a sophisticated 2-class classification model by employing XGBoost, which allowed it to detect fast system anomalies quickly and accurately. Sharma et al. \cite{sharma2018recommending} employed a two-stage classification approach to identify specialized software gurus by analyzing tweets. Their approach can be applied to identify specialized gurus in four domains (i.e., Android, JavaScript, Linux and Python) on a dataset of 86, 824 Twitter users for evaluating its effectiveness.

 \begin{framed}
  \begin{samepage}
\textit{
\textbf{Summary of answers to RQ3:}
\begin{enumerate}
	\item The application of predictive models in 22 research topics throughout the SDLC were summarized.
	\item Software testing and debugging is the domain in which 50\% of predictive models have been applied in the selected primary studies. 40 studies use predictive models for software defect prediction.
	\item Only 4 of the studies worked on requirements classification using predictive models. Researchers may want to focus more attention on using predictive models in analyzing software requirements and testing.
 \end{enumerate}
 }
 \end{samepage}
\end{framed}

%% file: Threats_to_Validity.tex
\section{Threats to Validity}

%It is crucial to clarify the potential threats that affect the outcomes of our research. The following are the validity threats identified during this survey.

\vspace{0.1cm}\noindent{\bf Publication Bias. }
Publication bias is the issue of publishing more positive results over negative ones. Claims to reject or support a hypothesis will be biased if the original publication is suffering from bias. A tendency toward positive study outcomes rather than others leads to biased and even possibly incorrect conclusions, while some preferences in publishing are useful. Studies with a null result might not always be worse than studies with significant positive results, although significant results have a statistically higher chance of getting published \cite{kitchenham2004evidence}. In this paper, we select six top and prevalent software engineering venues. Thus, to some extent, the publications included in these venues are high-quality, which can reduce or eliminate the influence of publication bias on conclusions.

\vspace{0.1cm}\noindent{\bf Search Terms. }
Finding all relevant primary studies is still a challenge to any survey or literature review-based study. To tackle this issue, a detailed search strategy was prepared and performed in our research. Search terms were constructed with different strings identified by checking titles and keywords from relevant publications already known to the authors. Alternative synonyms and spellings for search terms were then modified by consulting an expert. These procedures provided a high confidence that the majority of the key studies were identified.

\vspace{0.1cm}\noindent{\bf Study Selection Bias. }
The publication selection process was carried out in two stages. In the first stage, studies were excluded based on the title and abstract independently by two researchers with extensive experience in software engineering. We conducted a pilot study of publication selection process to place a foundation in order to better understand the inclusion/ exclusion criteria, and evaluated inter-rater reliability to mitigate the threat emerged from the researchers’ personal subjective judgment. When two researchers could not reach an agreement on a study, a third researcher was consulted. In the second stage, studies were eliminated based on the full paper. Thus there is little possibility to miss relevant studies through this well-established study selection process.

\vspace{0.1cm}\noindent{\bf Data Extraction. }
Data extraction was conducted by two of the researchers, and the data extracted from the relevant studies was rechecked by the other researchers. Disagreements in the data extraction process were discussed and solved after the pilot data extraction so that the researchers could complete the data extraction process following the refinement of the criteria, improving the validity of our analysis.

%% file: Challenges_and_opportunities.tex
\section{Challenges and opportunities}

A number of research challenges and opportunities relating to the use of predictive models in software engineering remain, requiring further investigation. In this section, we discuss the limitations of using predictive models in different SE research areas and conclude four key findings pertaining to challenges when building these models. We also present several recommendations for future work.

\subsection{Challenges}

\subsubsection{Challenge 1: Quality of Datasets}

Source datasets for training and testing are critcial for predictive model-based approaches. The most significant challenge for most studies is whether the information in the datasets available reflect ground truth about the related software engineering tasks. With different research areas, there are a large number of datasets that can be used for experiments. However, the datasets vary much in quality, including  bias, noise, size, imbalance and mislabelling. Any of these will influence the effectiveness of the predictive models used. For example, a dataset with much noise may cause underfitting in building predictive models. A dataset with small size may cause overfitting in building predictive models. A heavily imbalanced dataset may even fail to build useful models at all.

\textbf{Data imbalance:}
 The datasets are usually not distributed evenly \cite{jing2016improved}. In some cases, over 90\% of the total data have data imbalance problem. Some datasets even contain a large proportion of false positive instances. This issue causes the class imbalance problem, which has a large influence on the performance of predictive models. Learning a minority class is difficult when a classifier is trained on imbalanced data. Therefore, the classifier is skewed toward the majority class resulting in a lower rate of detection. Several studies proposed different methods to address this problem, such as instance re-weighting \cite{nagappan2005use}, data re-sampling \cite{mockus2009amassing}, and selective learning strategy \cite{hosseini2018benchmark}.

\textbf{Noise in datasets:}
Compared with the acceptable range of the feature spaces, data instances having excessive deviation can be referred to as outliers \cite{hosseini2017systematic}. Classifiers trained on noisy data with many outliers (e.g., incorrect, missing information) are expected to be less accurate. Since the problem of noisy data is inevitable from time to time when dealing with certain datasets such as software repositories, the usage of such noisy data is likely to pose a threat to the validity of many studies \cite{kim2011dealing}. Prior studies demonstrated that noisy may creep into datasets and impact predictive models when collecting data carelessly \cite{kim2011dealing}. There are two lines of related studies focusing on this problem. The first conducted a comprehensive investigation of the impact of noisy data on models performance \cite{kotsiantis2006data}. The second line of studies proposed different textual similarity approaches to address this challenge \cite{kotsiantis2006data}.

\textbf{Size of datasets:}
Most of selected studies mentioned the problem of generality of their proposed approaches as an external threat due to lack of enough data. Using representative datasets is one of crucial factors towards performing experiments \cite{rahman2013sample}. For instance, some studies trained their predictive models using source code in a single programming language such as C, Java or Python. This means the training classes cannot comprehensively cover all possible patterns and may prevent those algorithms from addressing corresponding tasks in all cases. In addition, overfitting may affect models' ability when training models without large-scale datasets. A number of studies have worked on alleviating this problem by constructing large-scale datasets from various software repositories \cite{liu2018androzooopen}. However, a sufficiently large scale volume of data in some communities is not yet available to be used.

\textbf{Data Privacy:}
Data privacy issues arise when the confidentiality of the data is a concern for the project owners. This issue in turn may cause the owners to not contribute to the pool of available data even though such data might contribute to further research efforts. Privacy is considered by Peters et al. \cite{peters2013balancing} in the context of Cross-Company Defect Prediction (CCDP). They obfuscate the data in order to hide the project details when they are shared among multiple parties.

\textbf{Data heterogeneity:}
The outcome of predictive models can be easy influenced by the similarity of source and target data distributions, and this problem is called data heterogeneity. As expressed by Canfora et al. \cite{canfora2015defects}, most software projects are heterogeneous and have various metric distributions. Moreover, certain context factors (e.g., size, domain, and programming language) have a huge impact on data heterogeneity. Machine learning techniques can work well under the assumption that source and target data have the same distribution. Some studies \cite{zimmermann2009cross, turhan2009relative} present different approaches such as filtering (F), transformation (DT), normalization (N) and feature matching to tackle this problem.

\subsubsection{Challenge 2: Feature selection}

As many candidate features are introduced to better represent various research domains, many studies extracted and used  a greater variety of features to train predictive models, increasing the size of research space of features. However, not all features used are beneficial to improving a predictive model's performance \cite{dam2018automatic}. First, some interdependent features do not fit when applied to build models. Another reason is that using too many features may result in overfitting. Therefore, how to select the most suitable features in a huge research space has become a critical challenge. Feature selection techniques ensure that useless features are removed. Some studies \cite{shivaji2012reducing, rahman2013and} employed such techniques in order to optimize performance and scalability. A good feature extractor should meet two main principles. First of all, it should be automatic so that it won’t cost too much manual effort. Second, it can reduce feature dimension while keeping feature quality.  That is, given a specific task, the feature extractor can identify and preserve relevant features and remove irrelevant features to improve the performance of predictive models. Automatic feature selection may even be possible in some research directions \cite{dam2018automatic}.

\subsubsection{Challenge 3: Evaluation metric selection}

Diverse evaluation metrics have been used to evaluate the effectiveness of different models. However, evaluation metrics may introduce bias. For example, \cite{huang2018identifying, zhou2018far} highlight the problem of the suitability of the chosen evaluation metrics. Most primary studies solve this challenge by utilizing standard metrics. Instead of using standard metrics such as precision, recall and F-measure, many studies \cite{lo2012active, fritz2014using, nagappan2005use, zhou2014will} have defined new evaluation measures to evaluate the performance of their proposed frameworks when addressing the same problem. This makes it difficult to evaluate the performance of their frameworks. On the other hand, the measures used for evaluation models keep changing thanks to new techniques being found. Therefore, it is a challenging task to achieve optimal selection of evaluation measures.

\subsubsection{Challenge 4: Guidelines for the selection of suitable predictive models}

From our survey, we see that there are various software engineering tasks leveraging predictive models. However, different predictive models fit better for different tasks. With various tasks and various predictive models, there is a need to have reliable guidelines for how to select a good predictive model for a specific tasks. Among the reviewed papers, some studies have proposed frameworks for specific tasks to make comparison of various predictive models \cite{ghotra2015revisiting, mori2019balancing, peters2017text, dejaeger2011data, mittas2012ranking}. However, almost all of them are based on experiments, which always have threats to validity. There ideally should be some theoretical guidelines that can help researchers select the right predictive models according to the characteristics of the specific SE tasks.

The selection of classifiers is another possible source of bias. Given the variety of available learning algorithms, there are still others that could have been considered. An appropriate selection might be guided by the aim of finding a meaningful balance ,or trade-off, between established techniques and novel approaches.

\subsection{Opportunities}

\subsubsection{Opportunity 1: Leveraging the power of big data}
The scale of available SE data has rapidly become larger and larger. A growing number of studies tend to be large-scale. The biggest advantage of big data is that it generally leads to robust predictive models, improving the practicability and generality. If a study only uses several small datasets to build predictive models it may not be generalisable and its results may have serious threats to validity. On the contrary, results achieved using big data are usually more convincing and more likely to have generality. In addition, deep learning techniques usually require a large amount of data to learn their predictive models, since there are many more parameters to learn than when using traditional predictive models. With the trend to using big data, reducing the time and space cost for data computation can also become a challenging issue.

\subsubsection{Opportunity 2: Neural network based predictive models}

Studies that apply predictive models to software engineering tasks occurred over several decades. Although researchers have achieved significant improvement from simple linear regression to complex ensemble learning, they have been bottlenecks for addressing many software engineering tasks. Recently, deep learning, as an advanced predictive model, has become more and more popular. Some studies that tried to leverage deep learning for software engineering tasks have achieved even better performance than all other start-of-the-art techniques. %However, till now the application of deep learning to software engineering tasks is limited in number.

The biggest advantage of deep learning is that it can automatically generate more expressive features that are better for learning predictive models, which can’t be done when using traditional predictive models \cite{dam2018automatic}. As mentioned above, many software engineering tasks face the feature extraction challenge, i.e., either it is hard to manually infer proper features or there are too many features needing to be carefully selected for building their predictive models. Therefore, leveraging deep learning for various software engineering tasks to improve their performance is a promising direction.

A well-developed predictive model often requires training with a great number of instances. However, there is not enough data to build such models in certain research domains. This problem can be solved by leveraging a novel deep learning technique referred to as meta-learning \cite{finn2017model} that simulates humans' behavior when learning new concepts and skills much faster and efficiently with many fewer data instances. So far, few studies in software engineering have adopted a good meta-learning model capable of well adapting or generalizing to new tasks and new environments that have never been encountered during training time.  Therefore, in addition to constructing more large scale datasets, meta-learning could be adopted as another significant solution for training well-performed machine learning models with a few training examples. This technique has been verified in other fields such as image classification and robots.

\subsubsection{Opportunity 3: Assessment and selection of predictive models}

Clearly, computational efficiency and transparency are desirable features of candidate classifiers and  appear to be a promising area for future research to formalize these concepts, e.g., by developing a multidimensional classifier assessment system \cite{lessmann2008benchmarking}.

Consequently, the assessment and selection of a predictive model should not be based on predictive accuracy alone but should be comprised of several additional criteria. These include computational efficiency, ease of use, and especially comprehensibility. Comprehensible models reveal the nature of the detected relationships and help improve our overall understanding of software failures and their sources, which, in turn, may enable the development of novel predictors in various research directions.

Efforts to design new software metrics and other explanatory variables appear to be a particularly promising area for future research.  They have the potential to achieve general accuracy improvements across all types of classifiers.

\subsubsection{Opportunity 4: predictive models in specific research domains}

\textbf{Predictive models in software requirements and testing:} According to our analysis we found that predictive models were rarely applied in certain domains, such as software requirements, software design and implementation. There are many different practical problems that are able to be tackled by leveraging predictive models in these two topics. We believe that researchers can try to apply state-of-the-art models on classification, prioritization and prediction tasks in both software requirements and testing.

%\textbf{Predictive models in code prediction analysis:}
\textbf{The usability analysis of prediction tasks:}
There is a significant difference between the accuracy and usefulness of predictions across projects. Projects with a very low percentage of positives and a very high number of classes are intuitively hard to predict for most approaches. However, this task is even harder for humans to perform manually. User studies are therefore needed to provide insights into when and where automated techniques are useful to humans, expecting to find a break even point where automation becomes more accurate than fully manual techniques.

% the usability analysis of prediction tasks:

Another potential direction is understanding the utility of change impact analysis. The ability to predict which classes are impacted by a new requirement can potentially support a range of tasks including refactoring decisions, defect prediction, and effort estimation \cite{falessi2018leveraging}. An increased understanding of such tasks could enable developers to build more effective prediction algorithms.

\textbf{Predictive models in identifying different defects:}
As mentioned above, a significant number of studies have applied predictive models to identify various types of defects, such as logical defects, syntax defects, interface defects, security defects, and performance defects. However, there are no studies that investigate on which defect types the existing defect predictive models work the best. In future work, researchers can conduct studies to investigate the performance of different categories of predictive models on predicting different types of defects. We may also add specific contexts when performing experiments.

%% file: Conclusion.tex
\section{Conclusion}

We wanted to analyse the use of predictive models in software engineering. A comprehensive analysis was performed to answer our defined research questions using a systematic review covering 145 selected papers published between 2009 and 2019 in top SE venues. These studies were identified by following a systematic series of steps and assessing the quality of the studies. Our key findings from this study are summarized below:

\begin{itemize}
\item The cumulative number of predictive model related studies shows an increasing trend over the last decade, and most of the selected primary studies focus on proposing novel approaches. %Over 90\% studies were published in conferences and journals.
\item We found 52 different predictive models were employed in software engineering tasks. These models can be classified into three categories -- base learners, ensemble learners and deep learners. 
\item Logistic Regression and Naive Bayes are the most widely used learning techniques to build predictive models for SE tasks to date. Several machine learning models are also popular models for addressing specific problems, including SVM and decision trees.
\item Recall, precision, and F-measure are the most frequently used performance metrics for evaluating the effectiveness of predictive models.
\item The application of deep learning techniques has seen an increasing trend in recent years. CNN and RNN are the two most popular deep learning models used in the studies for various SE tasks.
\item Most predictive models have been applied to the defect prediction task and 7 of 11 research topics where predictive models are frequently used belong to software maintenance and quality assessment domains.
\item We identified a number of challenges when using these predictive models for SE tasks, including issues with dataset quality, feature and evaluation metric selection, and lack of model selection guidelines. 
\end{itemize}

The main objective of this survey was to analyze  and  classify the use of existing predictive models and their related studies in SE to date. A range of future studies should include more applications of predictive models to requirements engineering and testing, better assessment and selection guidelines for different models for different SE tasks and use of meta-learning with neural network-based models. 

%, better data construction support,

%, in order to identify and classify predictive models used in software engineering, and give a summarization of research domains in which predictive models applied. Moreover, this survey aimed to conclude several challenges and potential opportunities, providing researchers with new directions for further studying.